\documentclass{article}

\usepackage{microtype}
\usepackage{graphicx}
\usepackage{subcaption}
\usepackage{booktabs} 

\usepackage{hyperref}

\usepackage[accepted]{icml2026}

\usepackage{amsmath}
\usepackage{amssymb}
\usepackage{mathtools}
\usepackage{amsthm}

\usepackage[capitalize,noabbrev]{cleveref}

\theoremstyle{plain}

\theoremstyle{definition}

\theoremstyle{remark}

\usepackage[textsize=tiny]{todonotes}
\usepackage{xspace}
\usepackage{multirow}
\def\name{\emph{CORRECT}\xspace}
\usepackage{subcaption} 
\usepackage[inline,shortlabels]{enumitem}
\newenvironment{denseitemize}{
\begin{itemize}[topsep=2pt, partopsep=0pt, leftmargin=1.5em]
  \setlength{\itemsep}{4pt}
  \setlength{\parskip}{0pt}
  \setlength{\parsep}{0pt}
}{\end{itemize}}

\usepackage{booktabs}   
\usepackage{multirow}   
\usepackage{makecell}   

\usepackage[most]{tcolorbox}
\usepackage{xcolor}
\usepackage{wrapfig}
\usepackage{booktabs}

\usepackage{xcolor}
\usepackage{verbatim}
\newcommand{\revision}[1]{{\color{blue}#1}}
\usepackage{algorithm}
\usepackage{algorithmic}

\icmltitlerunning{CORRECT: Condensed Error Recognition via Knowledge Transfer in Multi-agent Systems}
\tcbset{
  llmstyle/.style={
    enhanced,
    colback=#1!5,
    colframe=#1!70!black,
    fonttitle=\bfseries,
    coltitle=black,
    boxrule=0.6pt,
    arc=2pt,
    left=3mm, right=3mm, top=1mm, bottom=1mm
  }
}

\newtcolorbox{llmprompt}[1]{llmstyle=blue,title={#1}}

\newtcolorbox{llmredresponse}[1]{llmstyle=red,title={#1}}

\newtcolorbox{llmgreenresponse}[1]{llmstyle=green,title={#1}}

\newtcolorbox{llmprompt2}[1]{llmstyle=green,title={#1}}

\begin{document}

\twocolumn[
  \icmltitle{{CORRECT}: Condensed Error Recognition via Knowledge Transfer in Multi-agent Systems}



  \icmlsetsymbol{equal}{*}

  \begin{icmlauthorlist}
    \icmlauthor{Yifan Yu}{uiuc}
    \icmlauthor{Moyan Li}{amazon}
    \icmlauthor{Shaoyuan Xu}{amazon}
    \icmlauthor{Jinmiao Fu}{amazon}
    \icmlauthor{Xinhai Hou}{umich}
    \icmlauthor{Fan Lai}{uiuc}
    \icmlauthor{Bryan Wang}{amazon}
  \end{icmlauthorlist}

  \icmlaffiliation{uiuc}{University of Illinois Urbana-Champaign}
  \icmlaffiliation{amazon}{Amazon}
  \icmlaffiliation{umich}{University of Michigan}

  \icmlcorrespondingauthor{Fan Lai}{fanlai@illinois.edu}
  \icmlcorrespondingauthor{Bryan Wang}{ brywan@amazon.com}

  \icmlkeywords{Machine Learning, ICML}

  \vskip 0.3in
]



\printAffiliationsAndNotice{Part of this work was done while Yifan Yu was an intern at Amazon.} 

\begin{abstract}
 Multi-agent systems (MAS) are increasingly capable of tackling complex real-world tasks, yet their reliance on inter-agent coordination, tool use, and long-horizon reasoning makes error recognition particularly challenging. Minor errors can propagate across agents, escalating into task failures while producing long, intertwined execution trajectories that impose significant costs for both human developers and automated systems to debug and analyze. Our key insight is that, despite surface differences in failure trajectories (e.g., logs), MAS errors often recur with similar structural patterns. This paper presents \name, the first lightweight, training-free framework that leverages an online cache of distilled error schemata to recognize and transfer knowledge of failure structures across new requests. This cache-based reuse allows LLMs to perform targeted error localization at inference time, avoiding the need for expensive retraining while adapting to dynamic MAS deployments in subseconds. To support rigorous study in this domain, we also introduce \name-Error, a large-scale dataset of over 2{,}000 annotated trajectories collected through a novel error-injection pipeline guided by real-world distributions, and further validated through human evaluation. 
 Experiments across seven diverse MAS applications show that \textsc{CORRECT} improves error localization up to 19.8\% over existing advances while at near-zero overhead. Our code is publicly available at: \url{https://github.com/UIUC-MLSys/CORRECT}.
\end{abstract}

\section{Introduction}
\label{sec:introduction}
Multi-agent systems (MAS) have demonstrated success in various domains, including software development~\citep{qian2023chatdev,hong2024metagpt,zhang2024reactable}, scientific research~\citep{lu2024ai}, web navigation~\citep{zhou2023webarena}, and general-purpose task automation~\citep{wu2024autogen, intentnet-sigcomm}. By orchestrating multiple specialized agents, MAS can tackle challenges beyond the reach of single-agent systems (SAS) that rely on single LLMs for task solving.

However, as MAS increasingly scale in both complexity (e.g., sophisticated interactions with other agents and tools) and deployments, decisive error recognition that pinpoints the precise agent and step that first triggered the failure~\citep{zhang2025agent} has been a fundamental challenge~\cite{agent-prod-survey25}. Unlike SAS where failures can be traced to a single faulty output, MAS failures often emerge from cascading effects across agents: an error-prone step by one agent can propagate through downstream interactions, ultimately causing task failure~\citep{gao2025single}. Efficient decisive error recognition is essential for reliable MAS deployments, sustaining service quality, and guiding operational management such as safe agent upgrades, targeted restarts, and service monitoring~\citep{epperson2025interactive}.

Unfortunately, decisive error recognition in MAS remains open-ended due to three fundamental obstacles: (1) \emph{Generality}: MAS span a wide spectrum of applications, and even within an application, requests can exhibit drastically different error patterns, making it difficult to design methods that generalize. Existing advances~\citep{zhang2025agent} often resort to LLM-as-a-judge methods for generality, yet achieve $\leq$10\% accuracy. Recent efforts~\citep{ge2025introducing} to improve accuracy through fine-tuning not only hurt generalization, but fall short in (2) \emph{Data Efficiency}: obtaining labeled data in error recognition is notoriously expensive and inherently ambiguous. For example, in the \textsc{Who\&When} dataset~\citep{zhang2025agent}, annotators spent over 30 expert hours labeling fewer than 200 trajectories, yet disagreement rates exceeded 50\%. This makes any training-based approaches ineffective without voluminous data; and (3) \emph{Computation Efficiency}: even with sufficient data, tuning LLMs for error recognition may be impractical to catch up with deployments where new error types arise continuously and sporadically, e.g., new attacks in cloud AIOps~\citep{intentnet-sigcomm}.


In this paper, we first notice that failures in MAS tend to recur with similar structures across requests (\S\ref{subsec:motivations}). 
Because a MAS application often relies on the same role specifications, orchestration rules, tool APIs, or verification policies, diverse requests often funnel into common decision skeletons. 
Our real-world analysis of \textsc{Who\&When} dataset proposed in~\citet{zhang2025agent} shows that over 80\% of failure trajectories have at least one counterpart with $\geq$0.8 semantic similarity in error logs. This suggests that error knowledge can be systematically \emph{distilled, cached, and reused}. 
However, naive approaches, such as in-context learning (ICL) that insert prior trajectories as in-context exemplars, quickly break down: logs can exceed 32,000 tokens~\citep{yang2025qwen3,dubey2024llama}, contain low-entropy noise, and even underperform zero-shot baselines (\S\ref{subsec:motivations}). 


In this paper, we present \name (\underline{CO}ndensed e\underline{R}ror \underline{REC}ognition via knowledge \underline{T}ransfer), a novel framework that elevates \emph{decisive error recognition} to a first-class systems problem for practical and reliable MAS. \name automatically distills prior failures into compact, reusable error schemata that encode their core signatures, triggering contexts, and propagation patterns. At runtime, when a failure occurs, \name retrieves and instantiates the most relevant schemata to diagnose the new trajectory, enabling accurate, training-free error recognition in dynamic online environments. We summarize our contributions as follows:

\begin{denseitemize}

\item
\textbf{CORRECT: the first schema-guided detector.} 
We introduce \name, the first framework that systematically distills recurrent MAS failures into compact error schemata and reuses them for decisive error recognition. In contrast to LLM-as-a-judge approaches that sacrifice accuracy for generality, and fine-tuning approaches that incur high data and compute costs, \name transfers error knowledge across requests at inference time via schema retrieval. This design improves step-level localization accuracy by up to 20 points over prior work~\citep{zhang2025agent}, while remaining training-free and lightweight for large-scale online deployment.

\item 
\textbf{\name-Error: a large-scale, and high-fidelity dataset for MAS error recognition.} To address the lack of reliable evaluation data, we construct \name-Error, a benchmark of over 2,000 multi-agent trajectories with fine-grained, step-level error annotations. \name-Error is generated via a novel error-injection pipeline guided by real-world failure distributions, producing data that combines scalable coverage with the realism of natural MAS failures. Extensive human validation shows strong alignment between synthetic and expert-labeled errors. Beyond enabling rigorous evaluation of \name, \name-Error establishes a reusable and extensible benchmark for the community.

\end{denseitemize}

These contributions advance the state of the art in MAS error recognition and establish a foundation for building more reliable, interpretable, and scalable multi-agent systems. We will release the datasets and benchmarks.




\section{Background and Motivation}
\label{sec:background}

\subsection{Decisive Error in Multi-Agent Systems}
\label{subsec:def_decisive_error}

Task failures in MAS often arise from specific \emph{decisive errors} that, once committed, make successful task completion impossible. We formalize this notion following prior work~\citep{zhang2025agent}. 
Consider a MAS executing a trajectory $\tau = \{(a_1, s_1), (a_2, s_2), \dots, (a_T, s_T)\}$, where agent $a_i$ performs step $s_i$. The outcome of the trajectory is denoted by $\mathcal{R}(\tau) \in \{0, 1\}$, with $1$ for success and $0$ for failure.  
A step $(a_k, s_k)$ in a failed trajectory $\tau$ is a \textit{decisive error} if replacing it with a correct alternative $\tilde{s}_k$ would change the outcome to success. The earliest decisive error is
$
(a^*, s^*) = \min_{k \in \mathcal{D}(\tau)} k
$, 
where $\mathcal{D}(\tau) = \{\,k : \mathcal{R}(\tau) = 0 \wedge \mathcal{R}(\tau_{[s_k \rightarrow \tilde{s}_k]}) = 1 \,\}$, and $\tau_{[s_k \rightarrow \tilde{s}_k]}$ denotes the modified trajectory where step $s_k$ is replaced.


Intuitively, a decisive error is \textbf{the earliest step} whose correction flips a trajectory's outcome from failure to success. Identifying such errors is fundamental: Unlike coarse-grained error recognition, which merely flags failed trajectories, decisive error recognition localizes the precise \emph{agent} and \emph{step} that initiated failure. This fine-grained attribution enables targeted interventions, such as tuning role specifications, refining orchestration logic, or upgrading individual agents, without costly overhauls of the entire system.


\subsection{Motivations of \name}
\label{subsec:motivations}

Automated decisive error recognition in MAS has primarily followed two directions: LLM as a judge and fine-tuning specialized LLMs. Both exhibit fundamental limitations in accuracy, generality, and efficiency.

\paragraph{Limitations of Existing Advances.}
LLM-as-a-judge methods were initially designed to rate the quality of LLM outputs, achieving up to 80\% agreement with human preferences~\citep{zheng2023judging}. \citet{zhang2025agent} extended this paradigm to MAS error attribution, introducing three variants: (i)  \textit{all-at-once}, which provides the full error log to the LLM and asks it to identify the error step; (ii) \textit{step-by-step}, which incrementally reveals the trajectory and checks errors at each step; and (iii) \textit{binary search}, which recursively partitions the log to localize the error. However, as MAS becomes more complex, which elongates the failure trajectory, these methods lose diagnostic precision (Figure~\ref{fig:token_dist}): on the \textsc{Who\&When} dataset~\citep{zhang2025agent}, Qwen-2.5-7B achieves only 3.5\% step-level accuracy.

Fine-tuning-based approaches, whether supervised or reinforcement learning–based, face substantial efficiency and expense challenges. Their success hinges on large-scale, high-quality labeled datasets. Yet annotating MAS failures is prohibitively expensive: annotators must disentangle long, interdependent interactions across agents and tools, taking 30 expert hours for annotating fewer than 200 trajectories. Worse, error trajectories vary across applications and requests, making it infeasible to generalize across settings. 

\begin{figure}[t]
  \centering
  \hfill
  \begin{subfigure}[t]{0.48\linewidth}
    \centering
    \includegraphics[width=\linewidth]{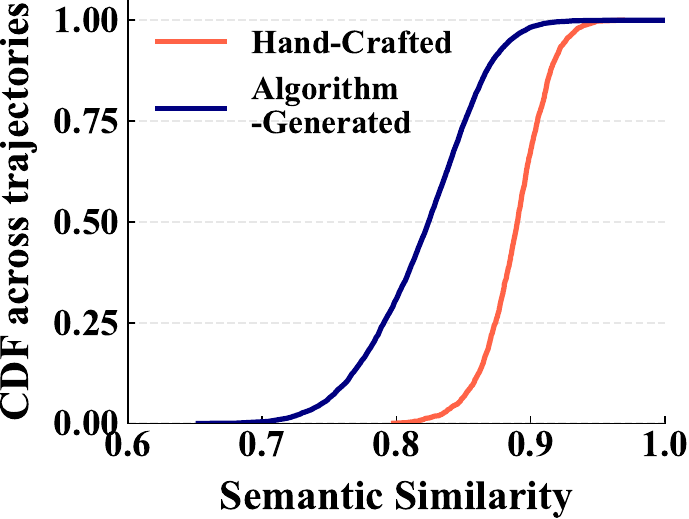}
    \caption{Failure trajectories exhibit high semantic similarities (\textsc{Who\&When} dataset).}
    \label{fig:similarity}
  \end{subfigure}
  \hfill
  \begin{subfigure}[t]{0.48\linewidth}
    \centering
    \includegraphics[width=\linewidth]{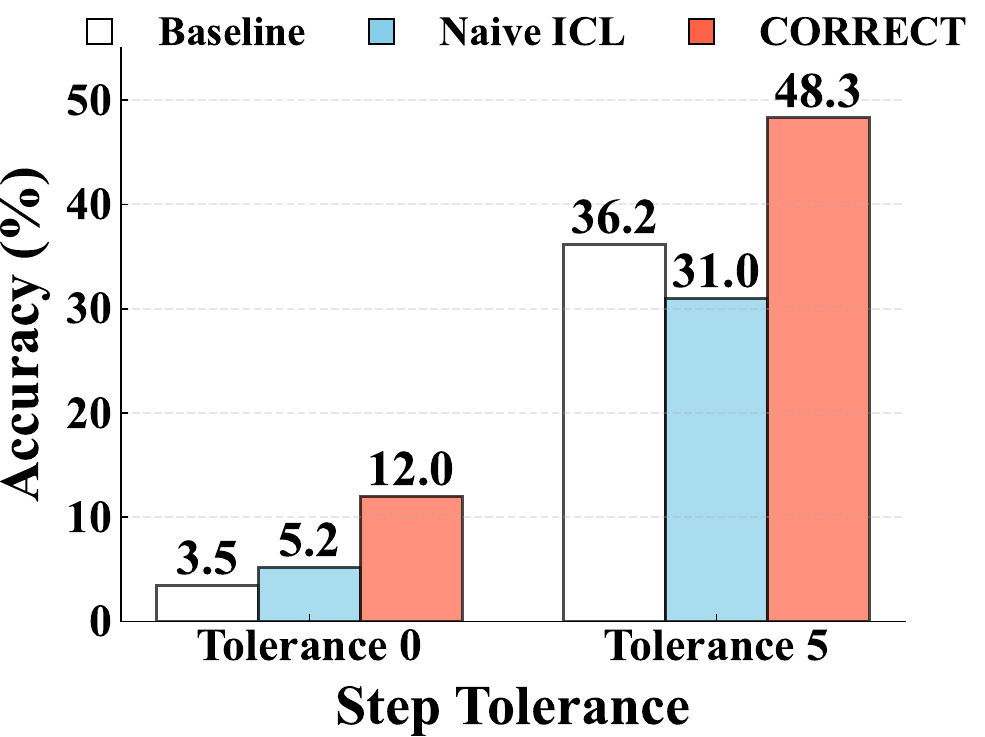}
    \caption{Performance of naive ICL and our method (\textsc{Who\&When} dataset).}
    \label{fig:naive+icl}
  \end{subfigure}
  \caption{Analysis of failure traces on the \textsc{Who\&When} dataset.}
  \vspace{-.4cm}
  \label{fig:analysis}
\end{figure}

\paragraph{Pervasive Error Similarity Yet Hard to Reuse.}
Despite these challenges, our analysis of the \textsc{Who\&When} dataset (Figure~\ref{fig:similarity}) reveals that more than 80\% of failed requests share a semantic (cosine) similarity above 0.8, measured via BERT-based embeddings. 
This reveals an underexploited opportunity: error knowledge recurs and could, in principle, be reused across requests. 
A natural attempt is to adopt in-context learning (ICL), retrieving and appending similar trajectories to guide error recognition. However, our experiments (Figure~\ref{fig:naive+icl}) show that such a strawman approach even degrades accuracy, due to (i) \textit{extreme trajectory length}: 17\% of trajectories exceed 32K tokens (details in Appendix~\ref{app:failure-length}), surpassing the context length of most LLMs (e.g., Qwen3~\citep{yang2025qwen3}); and (ii) \textit{low signal-to-noise ratio}: execution trajectories interleave request-specific details and tool outputs with the true error-inducing steps, diluting critical information.

\section{CORRECT: Condensed Error Recognition via Knowledge Transfer}
\label{sec:detect-pipeline}

Our observations call for a novel approach that can \emph{systematically reuse structural error knowledge} without overwhelming context or succumbing to noise, while remaining general, data-efficient, and lightweight for real-time MAS deployment.  To these ends, 
we introduce \name (\textbf{CO}ndensed e\textbf{R}ror \textbf{REC}ognition via knowledge \textbf{T}ransfer), the first framework that distills past errors into compact error schemas and adaptively applies them for accurate decisive error recognition, entirely without training.  We next introduce how CORRECT achieves this via three phases: (1) \emph{Offline schema extraction}, (2) \emph{Online schema-guided error recognition}, and (3) \emph{Dynamic schema management}. 




\subsection{Error Schema Extraction}
\label{sec:schema}

\begin{figure}[t]
    \centering
    \includegraphics[width=\linewidth]{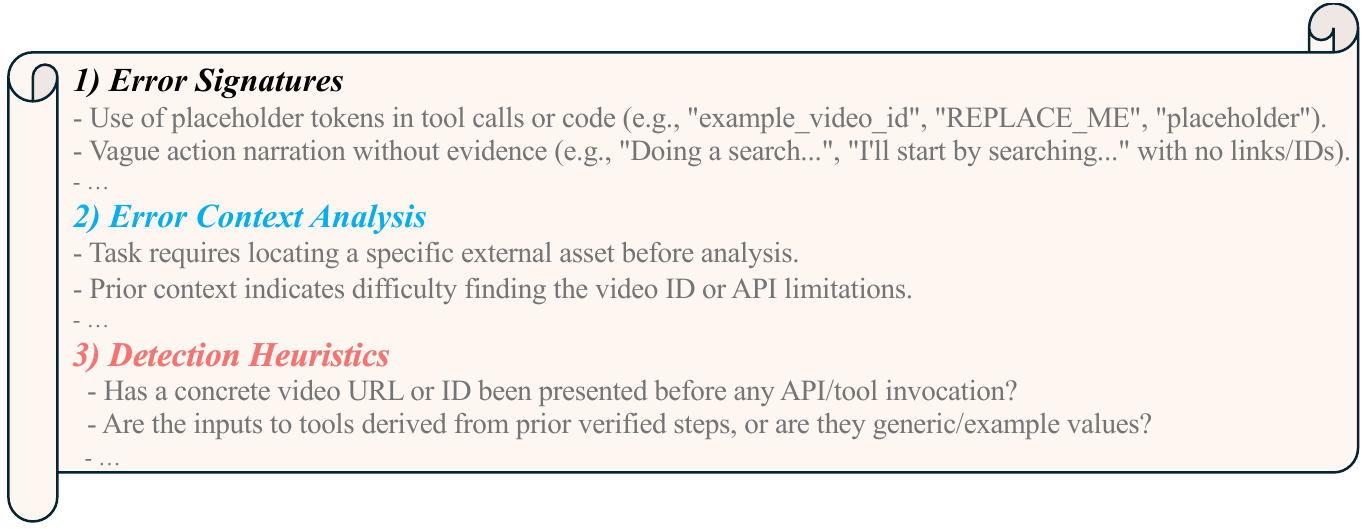}
    \caption{Example of an error schema generated on the \textsc{Who\&When} dataset.}
    \label{fig:sample_error_schema}
    \vspace{-0.3cm}
\end{figure}

Given an annotated error trajectory $\mathcal{T} = \{(a_i, s_i, r_i)\}_{i=1}^n$, where $a_i$ denotes the agent at step $i$, $s_i$ represents the step content, and $r_i$ is the corresponding result, along with the identified error at step $s_e$ and error reason $r_e$, \name generates an error schema $\mathcal{S}$ capturing (Figure~\ref{fig:sample_error_schema}): (1) \emph{Error Signatures} $\Sigma$: characteristic patterns such as agent actions, interaction sequences, and key behavioral markers; (2) \emph{Error Context Analysis} $\mathcal{C}$: detailed analysis of the conditions that led to the error, including agent states, task progress, and environmental factors; and (3) \emph{Detection Heuristics} $\mathcal{H}$: actionable rules and guidelines for identifying similar errors.

To minimize human efforts, \name leverages  LLMs (e.g., GPT-5 or Qwen3) to generate error schemas. We discuss how to ensure the quality of the schema in Section~\ref{sec:online-manage}. 


\paragraph{Clustered Schema Extraction.}
Even with LLMs, generating a schema for every trajectory can be costly, considering voluminous requests in practical MAS deployments. In fact, doing so is unnecessary because schema reuse often follows a long-tailed distribution: a small number of schemas are frequently reused, while most are rarely applied (\S\ref{sec:eval}). To exploit this, \name performs the following offline procedure: (1) \emph{Trajectory Clustering:} Failure trajectories are embedded semantically and clustered to group similar error patterns,  and then (2) \emph{Cluster-level Schema Generation:} One representative schema is generated per cluster, capturing the common error structure. 

The number of clusters is selected using a data-driven criterion that balances schema compactness and diagnostic coverage. 
Let $\{\mathbf{e}_i\}_{i=1}^N$ denote the semantic embeddings of $N$ failure trajectories, and let $\mathcal{C}(K)$ be the clustering result with $K$ clusters. 
For each trajectory $i$, we compute its silhouette coefficient:
\[
s(i) = \frac{b(i) - a(i)}{\max\{a(i), b(i)\}},
\]
where $a(i)$ is the average distance between $\mathbf{e}_i$ and other points in the same cluster, and $b(i)$ is the minimum average distance to points in any other cluster.
We define the average silhouette score as $
S(K) = \frac{1}{N} \sum_{i=1}^{N} s(i)
$. 
We increase $K$ until the marginal gain in cohesion saturates, i.e.,
$
S(K+1) - S(K) < \epsilon
$, for a small threshold $\epsilon$. As shown in our cache-size ablations (\S\ref{eval:ablation}),  a relatively small set of schemas (often a few hundred) already reaches an accuracy plateau, confirming the effectiveness of our design.

\subsection{Schema-Guided Error Recognition Online}
\label{sec:online-manage}


Once a failure request requires diagnosis, we start with its trajectory $\mathcal{T}_{\text{target}}$ and retrieve the top-k relevant schemas from the cache via semantic similarity search (e.g., cosine similarity of embeddings):  
$$\text{sim}(\mathcal{T}_{\text{target}}, \mathcal{S}_i) = \cos(\text{embed}(\mathcal{T}_{\text{target}}), \text{embed}(\mathcal{S}_i))$$

With retrieved schemas, we prompt the LLM to perform error recognition by instantiating the schemas in the context of the target trajectory. These schemas, together with the trajectory and lightweight adaptation instructions, are passed to an LLM for diagnosis. Formally, the LLM receives  $(\mathcal{T}_{\text{target}}, \{\mathcal{S}_j\}_{j=1}^k, \text{prompt}_{\text{detect}})$ as input prompt and produces the error recognition result:  
$$
\text{result} = \text{LLM}_{\text{detect}}\big(\mathcal{T}_{\text{target}}, \{\mathcal{S}_j\}_{j=1}^k, \text{prompt}_{\text{detect}}\big).
$$  
By leveraging schemas as condensed expert knowledge, this schema-guided inference directs the LLM’s attention toward salient failure patterns, avoiding the cost and noise of processing full historical trajectories.

\paragraph{Adaptation with Schema Expansion and Distillation.} 

\name maintains an effective schema cache through two complementary mechanisms.  
First, \emph{schema expansion}: when user feedback confirms successful recognition, \name leverages the ground truth label from the user to generate and cache a new error schema following~\ref{sec:schema}. Priority is given to trajectories with low similarity to existing schemas (i.e., $\text{sim}(\mathcal{T}_{\text{new}}, \mathcal{S}_i) < \delta$ for all cached schemas), ensuring the cache captures diverse error patterns rather than redundant ones. 
Second, \emph{schema distillation}: expansion alone may yield suboptimal quality, and frequently accessed error schemas (cache hits $>\theta_{\text{hot}}$) may benefit from further refinement. In such cases, \name generates multiple candidate schemas and replays them against prior trajectories to select the discriminative one with the highest accuracy. 
Together, expansion ensures coverage for novel errors online, while distillation preserves cache efficiency by retaining only high-quality, discriminative schemas. 

In practical deployments, each schema is typically reused across voluminous requests, so the overhead of both operations is largely amortized: Our evaluations show that tens of schemas already achieve strong accuracy for thousands of request errors (thus amortized overhead in generating one schema is below 1\%). Moreover, both designs achieve consistent improvements across settings (e.g., $>\theta_{\text{hot}}$), and with these two mechanisms in place, the schema cache adapts robustly over time under evolving workloads (\S\ref{eval:e2e}).

Algorithm~\ref{alg:overview} summarizes \name runtime detection. The offline stage builds an initial cache by distilling clean, representative schemas from annotated trajectories (Line~1-Line~7). At test time, the system retrieves the top-$k$ relevant schemas to guide the LLM detector toward the most likely error (Line~8-Line~12). The cache tracks how often each schema is used so it can update itself: new schemas are added when no good match exists, and frequently accessed schemas are refined when needed. This keeps the cache compact, accurate, and aligned with live error patterns.

\begin{algorithm}[t]
\caption{CORRECT Framework}
\label{alg:overview}
\begin{algorithmic}[1]
\small
\REQUIRE Annotated trajectories $\{(\mathcal{T}, s_e, r_e)\}$; target $\mathcal{T}_{\text{target}}$
\ENSURE Error recognition result $(a^*, s^*, c)$
\medskip
\STATE \textbf{Offline Schema Extraction} \revision{(Sec~\ref{sec:schema})}
\STATE Cluster trajectories by semantic similarity \label{algo:off-start}
\FOR{\textbf{each} annotated trajectory $(\mathcal{T}, s_e, r_e)$}
    \STATE $\mathcal{S} \gets \text{LLM}_{\text{extract}}(\mathcal{T}, s_e, r_e)$
    \STATE Apply filtering/distillation
    \STATE $\mathcal{C}.\text{put}(\mathcal{S})$ \label{algo:off-end}
\ENDFOR 
\medskip
\STATE \textbf{Online Schema-Guided Error Recognition} \revision{(Sec~\ref{sec:online-manage})}
\STATE $\mathbf{e} \gets \text{embed}(\mathcal{T}_{\text{target}})$
\STATE $\{\mathcal{S}_j\}_{j=1}^k \gets \mathcal{C}.\text{search\_top\_k}(\mathbf{e})$
\STATE $\forall j: \mathcal{C}.\text{update\_access}(\mathcal{S}_j)$
\STATE $(a^*, s^*, c) \gets \text{LLM}_{\text{detect}}(\mathcal{T}_{\text{target}}, \{\mathcal{S}_j\})$
\medskip
\STATE \textbf{Dynamic Schema Management} \revision{(Sec~\ref{sec:online-manage})}
\IF{user feedback confirms successful recognition}
    \IF{$\text{sim}(\mathcal{T}_{\text{target}}, \mathcal{S}_i) < \delta$ for all $\mathcal{S}_i \in \mathcal{C}$}
        \STATE Extract ground truth label from data
        \STATE Distill new schema $\mathcal{S}_{\text{new}}$
        \STATE $\mathcal{C}.\text{put}(\mathcal{S}_{\text{new}})$
    \ENDIF
\ENDIF
\IF{$\mathcal{C}.\text{access\_count}(\mathcal{S}_j) > \theta_{\text{hot}}$}
    \STATE $\{\mathcal{S}_i'\}_{i=1}^m \gets \text{LLM}_{\text{extract}}^m(\mathcal{T}_j, s_{e,j}, r_{e,j})$
    \STATE Evaluate by replaying on prior trajectories
    \STATE $\mathcal{S}^* \gets \arg\max_{\mathcal{S}_i'} \text{accuracy}(\mathcal{S}_i')$
    \STATE $\mathcal{C}.\text{replace}(\mathcal{S}_j, \mathcal{S}^*)$
\ENDIF
\STATE \textbf{return} $(a^*, s^*, c)$
\end{algorithmic}
\end{algorithm}

\vspace{-0.2cm}

\section{\name-Error: A Large-Scale Error Detection Benchmark}
\label{sec:dataset}

Existing efforts for trajectory-level error analysis are limited in both scale and diversity, and human annotation is costly and difficult to scale (\S\ref{subsec:motivations}). To bridge this gap and evaluate \name's effectiveness (\S\ref{sec:detect-pipeline}), we introduce \name-Error, a large-scale benchmark that faithfully reflects the distribution of natural errors in real-world MAS. 

\subsection{Bootstrap Error Synthesis Pipeline}
\label{subsec:dataset_creation}

In building \name-Error, we develop a novel bootstrap methodology that uses a small set of human-annotated error trajectories as seeds for scalable error generation, blending realism with controllability. It follows a three-stage pipeline that distills human expertise into scalable synthetic data while preserving the structural and semantic integrity of real-world error patterns.

\paragraph{Stage 1: Diverse Trajectory Collection.}  
We first generate a large corpus of successful multi-agent trajectories spanning heterogeneous tasks and domains. In parallel, we curate a smaller but high-quality set of human-annotated error trajectories. These human-labeled examples serve as reference exemplars of realistic failure dynamics, capturing both localized mistakes and their downstream propagation.

\paragraph{Stage 2: Semantic Error Schema Matching.}  
Each successful trajectory is paired with its closest human-labeled error trajectory using semantic similarity measures that account for both high-level task goals and fine-grained agent interactions. This alignment ensures that the selected error schema is contextually aligned with the target trajectory, avoiding unrealistic mismatches. Then we use LLMs (e.g., GPT-5) to devise an error injection strategy that specifies (i) where in the target trajectory to introduce the error, and (ii) how to adapt the error pattern while preserving its semantics. 

\paragraph{Stage 3: Contextual Error Injection.}  
Following the injection strategy, we prompt GPT-5 that generated the original successful trajectory to introduce an erroneous action at the designated point. This guarantees consistency in linguistic and behavioral style while embedding a realistic failure. 
We provide the prompt used to generate natural, schema-consistent error
injections in Appendix~\ref{app:error_injection_prompt}.

\subsection{Human-Alignment Analysis}

\begin{figure}[t]
    \centering
    \includegraphics[width=0.9\linewidth]{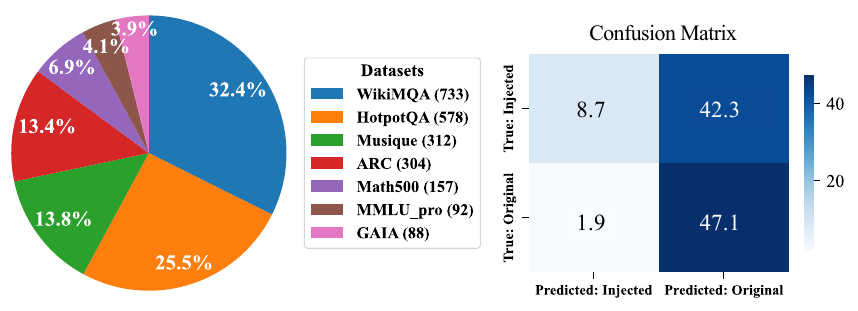}
    \caption{\name-Error includes diverse tasks. The synthesized data preserves high realism, where human labelers frequently misclassified synthetic errors as genuine ones.}
    \label{fig:dataset_distribution}
\end{figure}

Following our bootstrap pipeline (\S\ref{subsec:dataset_creation}), we synthesized over 2,000 trajectories across seven datasets (Figure~\ref{fig:dataset_distribution}), yielding \textbf{12.3$\times$} more data than \textsc{Who\&When} with cost over 3 billion tokens using GPT-5 series models and GPT-4o series models based on Magnetic-One~\citep{fourney2024magentic} and AutoGen~\citep{wu2024autogen}. The resulting benchmark spans diverse tasks, including multi-hop QA, common planning, mathematical reasoning, and scientific problem-solving. By leveraging limited human annotations as seeds, our novel pipeline generates diverse error scenarios at scale. 

%
%


To rigorously evaluate the authenticity of our synthesized data, we conducted a human evaluation study with four independent expert labelers, totaling over 120 hours of annotation effort. Each labeler was presented with an equal mix of synthetically injected and human-annotated error trajectories, without being informed of their origin. As shown in Figure~\ref{fig:dataset_distribution}, labelers struggled to distinguish between the two sources: 47.1\% of synthetic trajectories were misclassified as human-labeled, while only 42.3\% of genuine trajectories were correctly identified. This near-random classification performance, close to 50\% in both cases, indicates that synthetic errors produced by our pipeline are effectively indistinguishable from real-world failures.

Beyond individual judgments, we observe strong inter-annotator agreement on perceived authenticity. In particular, 94.4\% of synthetic trajectories were judged as genuine by at least two out of five labelers, and 52.9\% received unanimous consensus. These results suggest that our injected errors consistently exhibit the structural and semantic characteristics of natural MAS failures, which also reinforces the effectiveness of error schema design in \name. We provide a detailed human-alignment analysis, including confusion matrices and agreement statistics, in Appendix~\ref{app:alignment}.

Taken together, these findings validate that our error injection pipeline faithfully captures the nuanced failure patterns observed in real multi-agent systems. This methodology enables data generation that is \textit{cheap} (fully automated and low-cost), \textit{abundant} (scalable to millions of trajectories), \textit{precise} (with unambiguous ground-truth error labels), and \textit{realistic} (closely matching human-annotated error distributions), thereby providing a strong foundation for developing and evaluating effective MAS error recognition methods.

\section{Experiments}
\label{sec:eval}

We demonstrate that \name achieves significant accuracy improvements (up to 20\%) on \textsc{Who\&When} and an average gain of 28.7\% across MAS tasks (\S\ref{eval:e2e}) on \textsc{Correct-Error}, all at zero training costs. \name remains robust under distribution shifts arising from model updates, dataset variations, schema cache size, and the number of stored schemata (\S\ref{eval:ablation}). We provide case-studies with illustrative examples and detailed analysis in Appendix~\ref{app:multi_error}. 

\begin{table*}[t]
\centering
\tiny
\resizebox{0.85\textwidth}{!}{
\begin{tabular}{llcccc}
\toprule
\multirow{2}{*}{Method} & \multirow{2}{*}{Model} & \multicolumn{2}{c}{Human-Crafted} & \multicolumn{2}{c}{Algorithm-Generated} \\
\cmidrule(lr){3-4} \cmidrule(lr){5-6}
& & Acc@0 & Acc@1 & Acc@0 & Acc@1 \\
\midrule
\multirow{10}{*}{LLM-as-a-Judge} & Qwen-2.5-7b & 3.5 & 8.6  & 19.1 & 42.9 \\
 & Qwen-3-30b  & 1.7 & 5.2  & 15.1 & 42.9 \\
 & Qwen-3-80b  & 6.9 & 8.6  & 21.4 & 47.6 \\
 & Llama-8b    & 1.7 & 3.5  & 3.2   & 15.9   \\
 & DeepSeek-R1 & 3.5 & 17.2 & 23.8 & 54.0 \\
 & Gemini-2.5-flash & 5.2 & 13.8 & 31.8 & 56.0 \\
 & Gemini-2.5-pro   & 5.2  & 12.1  & 25.4 & 50.0 \\
 & GPT-4o-mini  & 3.5 & 12.5 & 12.7 & 39.7 \\
 & GPT-4o       & 3.5 & 10.3 & 18.3 & 50.0 \\
 & GPT-5-nano   & 1.7 & 12.1  & 19.1 & 41.3 \\
 & GPT-5        & 8.6 & 24.1 & 18.3 & 56.4 \\
\midrule
Fine-tuned LLM & Qwen-2.5-7b & 3.5 & 11.9 & 18.9 & 42.9 \\
Naive ICL     & Qwen-2.5-7b & 5.2 & 10.3 & 15.9 & 40.5 \\
MIPRO & Qwen-2.5-7b& 8.6 & 12.1 & 15.1 & 41.3 \\
\midrule
\multirow{5}{*}{\name} & Qwen-2.5-7b          & 12.1 (+8.6)  & 15.5 (+6.9)   & 19.8 (+0.7)  & 46.8 (+3.9) \\
 & Gemini-2.5-flash & 10.3 (+5.1) & 20.7 (+6.9)  & \textbf{38.9 (+7.1) } & 55.2 (-0.8) \\
 & Gemini-2.5-pro   & 6.9 (+1.7)  & 17.2 (+5.5)   & 24.6 (-0.8)  & 52.4 (+2.4) \\
 & GPT-5-nano       & 6.9 (+5.2)  & 17.2 (+5.0)  & 24.6 (+5.5)  & 44.4 (+3.1) \\
 & GPT-5            & \textbf{17.2 (+8.6) }&\textbf{ 32.3 (+8.2)}  & 38.1 (+19.8) & \textbf{58.8 (+2.4)} \\
\bottomrule
\end{tabular}
}
\caption{\name achieves higher error recognition accuracy over existing advances (\textsc{Who\&When} dataset).
Acc@0 denotes exact-step accuracy (the model must pinpoint the precise error step).
Acc@k denotes tolerant accuracy (a prediction is correct if it falls within $\pm k$ steps of the ground truth).
For rows corresponding to \name, ``(+X)'' implies its relative improvements over LLM as a judge.}
\vspace{-.4cm}
\label{tab:who_when_results}
\end{table*}

\paragraph{Models and Tasks.} 
We evaluate \name on both the human-annotated \textsc{Who\&When} benchmark and our high-quality benchmark, \textsc{Correct-Error}, which spans diverse tasks including multi-hop QA (HotpotQA~\citep{yang2018hotpotqa}, Musique~\citep{trivedi2022musique}, WikiMQA~\citep{xanh2020_2wikimultihop}), scientific reasoning (ARC~\citep{allenai:arc}, MMLU-Pro~\citep{wang2024mmlu}), mathematical reasoning (Math500~\citep{lightman2023lets}), and general agentic tasks including planning (GAIA~\citep{mialon2023gaia}).
\textsc{Who\&When} consists of two subsets: a Human-Crafted subset and an Algorithm-Generated subset. Experiments are conducted on both open- and closed-source models, including the Qwen~\citep{Yang2024Qwen25TR,yang2025qwen3}, Llama~\citep{dubey2024llama}, GPT~\citep{hurst2024gpt,openai2025gpt5}, DeepSeek-R1~\citep{guo2025deepseek}, and Gemini series~\citep{comanici2025gemini}. We mask each trajectory itself and avoid receiving its own error schema for preventing data leakage. Additional experimental details are provided in Appendix~\ref{sec:app_exp_settings}, including SFT details and hyperparameters. 

\paragraph{Baselines.}  
We compare against four advances:
\begin{denseitemize}
\item \emph{LLM-as-a-Judge}: a zero-shot prompting strategy where an LLM directly inspects trajectories without auxiliary guidance~\citep{zhang2025agent,peng2023instruction}; 

\item  \emph{Fine-tuning}: Qwen-2.5-7b-Instruct is trained on the full trajectory dataset to learn domain-specific failure patterns~\citep{ chen2025step,fu2025srft};   

\item  \emph{Naive In-Context Learning}, which inserts complete error trajectories as few-shot exemplars~\citep{10.1145/3731569.3764829}.

\item \emph{MIPRO}: optimizes prompts via Bayesian search over instruction–example combinations using success signals from prior runs~\citep{opsahl2024optimizing}.

\end{denseitemize}

\paragraph{Metrics.}
Following existing advances~\citep{zhang2025agent}, we report \emph{step-level accuracy}, which provides actionable debugging signals. To account for the ambiguity of error attribution, we additionally report accuracy@k, where predictions within $k$ steps of the ground truth are treated as correct, e.g., Acc@0 requires identifying the exact erroneous step, while Acc@1 tolerates an offset of one step. This better reflects practical debugging scenarios, where approximate localization is often sufficient. 

We report median performance over five independent runs.

\subsection{End-to-End Performance}
\label{eval:e2e}

\begin{table*}[t]
\centering
\tiny
\resizebox{0.9\textwidth}{!}{
\begin{tabular}{llcccccccc}
\hline
\multirow{2}{*}{Method} & \multirow{2}{*}{Tolerance} & \multicolumn{7}{c}{Dataset} & \multirow{2}{*}{\makecell{Avg.\\ Improv.}} \\
\cline{3-9}
& & Gaia & HotpotQA & Musique & WikiMQA & Arc & Math500 & MMLU-Pro & \\
\hline
\multicolumn{10}{l}{Synthesized by \textbf{GPT-4o-mini}} \\
\hline
\multirow{3}{*}{LLM-as-a-Judge} & Acc@1 & 28.6 & 34.8 & 27.8 & 14.7 & 64.0 & 10.2 & 58.3 & - \\
& Acc@3 & 42.9 & 59.4 & 77.8 & 55.9 & 75.0 & 23.4 & 62.5 & - \\
& Acc@5 & 50.0 & 63.8 & 77.8 & 64.7 & 78.0 & 35.6 & 66.7 & - \\
\hline
\multirow{3}{*}{CORRECT} & Acc@1 & 28.6 & \textbf{60.9} & \textbf{38.9} & \textbf{44.1} & \textbf{80.4} & \textbf{57.1} & \textbf{69.1} & \textbf{+20.1} \\
& Acc@3 & \textbf{50.0} & \textbf{94.2} & \textbf{88.9} & \textbf{88.2} & \textbf{88.2} & \textbf{87.8} & \textbf{88.2} & \textbf{+27.6} \\
& Acc@5 & \textbf{64.3} & \textbf{95.7} & \textbf{88.9} & \textbf{94.1} & \textbf{91.2} & \textbf{95.9} & \textbf{94.1} & \textbf{+28.7} \\
\hline
\multicolumn{10}{l}{Synthesized by \textbf{GPT-5-Nano}} \\
\hline
\multirow{3}{*}{Baseline} & Acc@1 & 16.7 & 14.7 & 11.9 & 6.44 & 62.8 & 41.8 & 50.0 & - \\
& Acc@3 & 27.8 & 48.9 & 43.9 & 35.6 & 69.6 & 57.1 & 64.7 & - \\
& Acc@5 & 38.9 & 65.8 & 58.8 & 56.1 & 71.1 & 64.3 & 70.59 & - \\
\hline
\multirow{3}{*}{CORRECT} & Acc@1 & \textbf{30.6} & \textbf{35.8} & \textbf{32.7} & \textbf{16.9} & \textbf{80.4} & \textbf{57.1} & \textbf{69.1} & \textbf{+16.8} \\
& Acc@3 & \textbf{44.4} & \textbf{72.7} & \textbf{61.2} & \textbf{49.9} & \textbf{88.2} & \textbf{87.8} & \textbf{88.2} & \textbf{+20.1} \\
& Acc@5 & \textbf{52.8} & \textbf{84.3} & \textbf{77.2} & \textbf{69.0} & \textbf{91.2} & \textbf{95.9} & \textbf{94.1} & \textbf{+18.9} \\
\hline
\end{tabular}
}
\caption{Performance comparison across multiple datasets. All numbers report the error recognition accuracy.}
\label{tab:combined_results}
\end{table*}

\paragraph{CORRECT achieves significant gains in error recognition accuracy (\textsc{Who\&When} dataset).}  
Table~\ref{tab:who_when_results} shows that \name consistently surpasses existing advances across both human-crafted and algorithm-generated subsets.  
On human-crafted data, \name raises Qwen-2.5-7B's exact-step accuracy from 3.5\% to 12.1\% (a 3.5$\times$ improvement), and improves GPT-5 from 8.6\% to 17.2\%. These gains extend to tolerant metrics as well, with GPT-5 + \name achieving 32.3\% at Acc@1 versus 24.1\% for the baseline. By contrast, fine-tuning (3.5\%) and naive ICL (5.2\%) offer only marginal improvements, suggesting that standard supervised learning and raw in-context trajectories fail to capture complex error patterns.  
On algorithm-generated data, \name maintains clear advantages, with Gemini-2.5-Flash improving from 31.8\% to 38.9\% (+7.1 points) and GPT-5 exhibiting the largest gain (+19.8 points).  
These consistent improvements across model families (Qwen, Gemini, GPT) and scales (7B to GPT-5) highlight the effectiveness of condensed error schemas.




\textbf{CORRECT delivers 17--28\% average improvements (\textsc{Correct-Error} benchmark).}  
Table~\ref{tab:combined_results} highlights \name's strong generalization across seven datasets. For GPT-4o-mini subset, \name improves average accuracy by 20.1\%, 27.6\%, and 28.7\% at Acc@1, Acc@3, and Acc@5, respectively using Qwen-2.5-7b. Gains are especially pronounced on knowledge-intensive tasks: HotpotQA (+26.1 points), WikiMQA (+29.4 points), and Math500 (+46.9 points). At higher tolerances, performance gaps widen further: \name reaches 94.2\%, 91.2\%, and 95.9\% at Acc@5, compared to baseline scores of 63.8\%, 78.0\%, and 35.6\%. \name exhibits a similar trend in GPT-5-nano subset, with average improvements of 16.8\%, 20.1\%, and 18.9\% across tolerance levels using Qwen-2.5-7b. Even on the challenging GAIA benchmark, \name achieves superior scores at Acc@5.

\begin{figure}[t]
  \centering
  \begin{subfigure}[t]{0.48\linewidth}
    \centering
    \includegraphics[width=\linewidth]{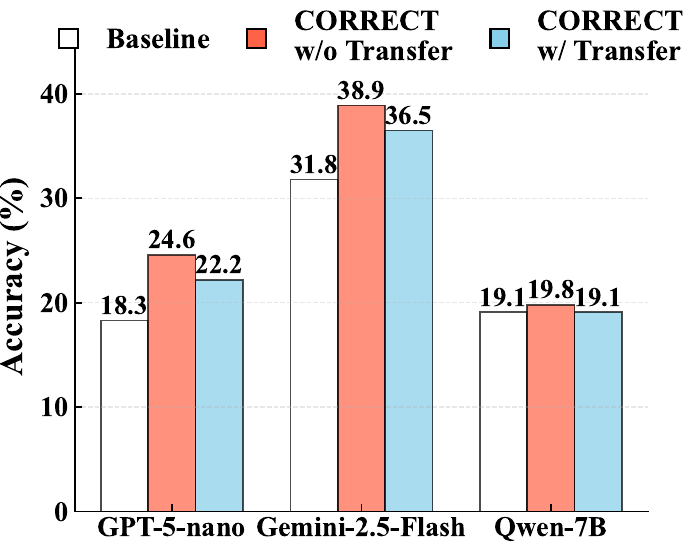}
    \caption{\name delivers improvements on Algorithm-Generated, with schemata from Hand-Crafted dataset.}
    \label{fig:ablation_transfer_dataset}
  \end{subfigure}
  \hfill
  \begin{subfigure}[t]{0.48\linewidth}
    \centering
    \includegraphics[width=\linewidth]{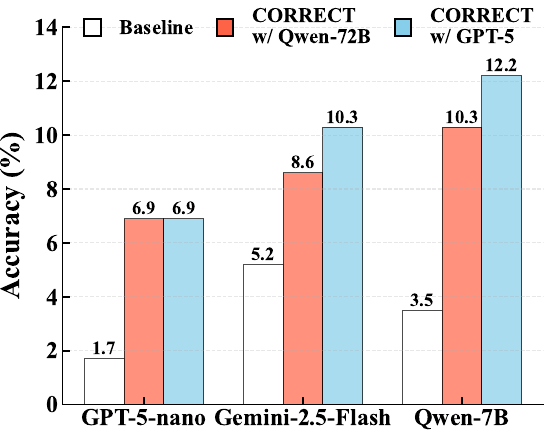}
    \caption{\name can adaptively upgrade its performance on the Hand-Crafted dataset with model upgrade.}
    \label{fig:model-upgrade}
  \end{subfigure}
  \hfill
  \caption{Ablation studies on transfer and model upgrade.}
  \label{fig:ablation_row}
  \vspace{-0.2cm}
\end{figure}

\textbf{Strong Schema Transferability across Datasets and Models.}
Figure~\ref{fig:ablation_transfer_dataset} shows that schemas distilled from human-crafted trajectories transfer effectively to algorithm-generated data. Across GPT-5-nano, Gemini-2.5-Flash, and Qwen-7B, \name with transferred schemas consistently outperforms baselines, with Gemini-2.5-Flash improving 31.8--36.5\%. This cross-domain transferability indicates that distilled schemas capture fundamental error patterns.  

Moreover, Figure~\ref{fig:model-upgrade} demonstrates that \name benefits directly from model upgrades: using GPT-5 instead of Qwen-72B as the schema generator raises detection accuracy from 8.6\% to 10.3\% for Gemini-2.5-Flash and from 10.3\% to 12.2\% for Qwen-2.5-7B. These adaptive gains confirm that better models yield higher-quality schemas that immediately enhance downstream performance. 






\paragraph{\name improves performance in online deployments.}
Real deployments often face continuously evolving request distributions. To
evaluate this setting, we initialize CORRECT using schemata distilled from
\textsc{Who\&When} and gradually augment the cache with schemata derived from incoming
HotpotQA and WikiMQA trajectories. As shown in
Table~\ref{tab:streaming_main}, \name maintains stable accuracy even when only
10--20\% of the newly observed schemata are incorporated. This reflects both
the strong transferability of our schemata and the small amount of
on-task data required for CORRECT to adapt effectively in online environments.
\begin{table}[h]
\centering
\footnotesize
\begin{tabular}{lcccc}
\toprule
Dataset & 0--10\% & 0--20\% & 0--50\% & 0--100\% \\
\midrule
Hotpot   & 35.8 & 40.6 & 41.2 & 39.4 \\
WikiMQA  & 70.8 & 68.6 & 67.6 & 69.1 \\
\bottomrule
\end{tabular}
\caption{
 Streaming-style adaptation: accuracy as more schemata from new domains arrive.
}
\label{tab:streaming_main}
\end{table}

\vspace{-0.2cm}
\subsection{Ablation Studies}
\label{eval:ablation}

\begin{figure}[t]
\centering
\hfill
\begin{subfigure}[t]{0.48\linewidth}
    \centering
    \includegraphics[width=\linewidth]{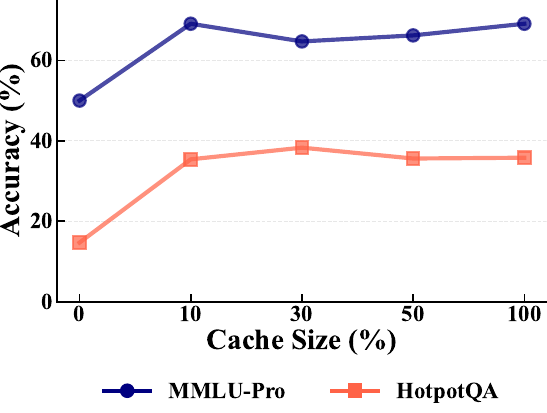}
    \caption{\name delivers robust improvements under different cache sizes.}
    \label{fig:ablation_cache_choice}
  \end{subfigure}
\hfill
 \begin{subfigure}[t]{0.48\linewidth}
    \centering
    \includegraphics[width=\linewidth]{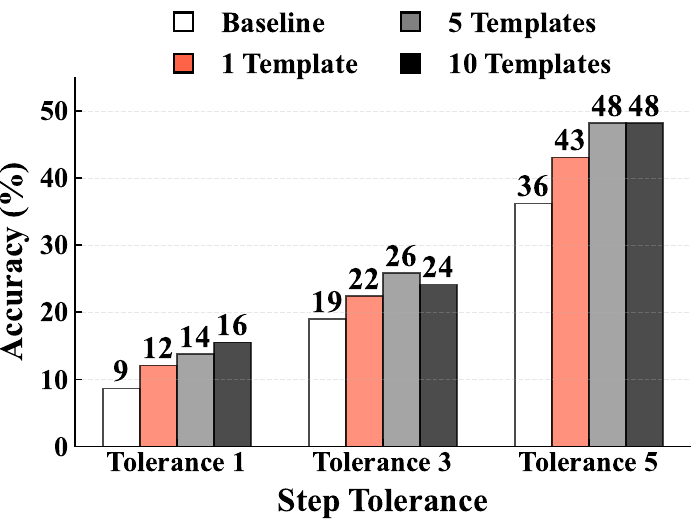}
    \caption{Performance of \name with different number of error schemata on Handcrafted subset of \textsc{Who\&When}.}
    \label{fig:abalation_num_schema}
  \end{subfigure}
\caption{Ablation studies on cache size and error schemata.}
\end{figure}

\paragraph{Impact of Schema Repository Size.}
Figure~\ref{fig:ablation_cache_choice} shows \name's robustness to cold-start scenarios and varying cache sizes. On \textsc{Correct-Error}, even with only 10\% of the schema library, \name achieves 69.1\% and 35.4\% Acc@1 on MMLU\_pro and HotpotQA, substantially outperforming baselines. Performance improves steadily with larger caches but plateaus beyond 50\% (tens of schemas), suggesting that a relatively small set of diverse schemas captures most common error patterns. This logarithmic growth pattern validates our clustering-based extraction strategy.


\paragraph{Impact of Number of Schemas in Online Error Recognition.}
Figure~\ref{fig:abalation_num_schema} demonstrates that retrieving and using a single schema already greatly improves baseline Acc@1 (12.1\% vs.\ 8.6\%). Accuracy increases as more schemas are added (13.8\% with 5 schemas, 15.5\% with 10), though gains diminish. At higher tolerances (Acc@5), settings converge to $\sim$48\%. These results show that a small set of well-matched schemas already efficiently captures most critical patterns, while additional schemas offer limited value.  

\begin{figure}[t]
  \centering
  \hfill
  \begin{subfigure}[t]{0.48\linewidth}
    \centering
    \includegraphics[width=\linewidth]{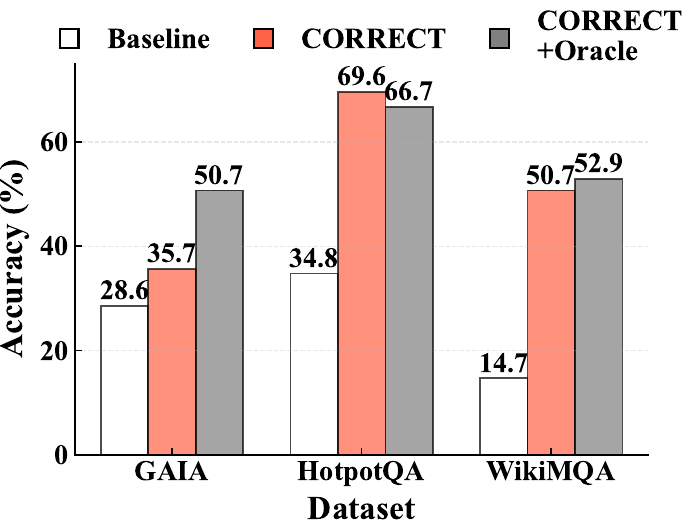}
    \caption{Performance of \name and \name with the variation using oracle error schema.}
    \label{fig:abalation_oracle}
  \end{subfigure}
  \hfill
  \begin{subfigure}[t]{0.48\linewidth}
    \centering
    \includegraphics[width=\linewidth]{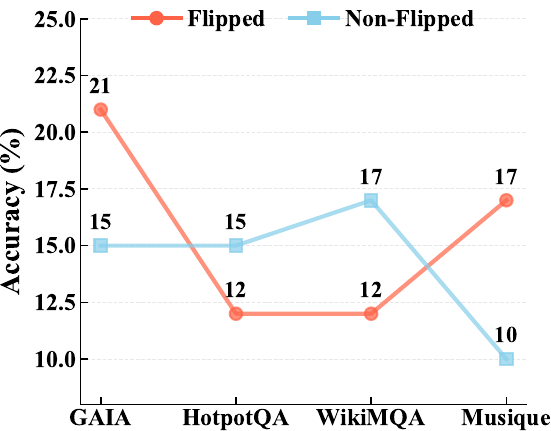}
    \caption{LLMs have low performance in recognizing the errors when they encounter them.}
    \label{fig:abalation_recognize_error}
  \end{subfigure}

  \caption{Studies on oracle guidance and error recognition.}
  \label{fig:ablation_error}
\end{figure}

\paragraph{Comparison with Oracle Error Schema.}
We compare \name against an oracle configuration where each trajectory uses its own ground-truth schema. As shown in Figure~\ref{fig:abalation_oracle}, the oracle achieves superior step-level accuracy, while \name with 5 retrieved schemas reaches over 71.5\% of oracle performance. This narrow gap shows that our semantic retrieval strategy effectively identifies schemas encoding near-equivalent knowledge to trajectory-specific patterns, validating that diverse MAS error patterns often share structural regularities.



\paragraph{LLMs can hardly Recognize Their  Errors During Execution.}
Figure~\ref{fig:abalation_recognize_error} shows that LLMs have limited metacognitive ability to detect their own errors during execution. When asked to identify injected errors at the exact step, they achieve only 21\% accuracy on \emph{flipped} trajectories (where errors alter the final answer) and 17–18\% on \emph{non-flipped} trajectories (where errors do not affect the outcome). This reveals a fundamental limitation: \emph{agents lack the self-awareness to recognize their own mistakes, regardless of downstream task success.} This motivates the need for external error-detection mechanisms like \name.

\paragraph{\name Introduces Negligible Overhead.}
Our GPT-5 evaluations show that API usage increases only slightly (\$0.86 $\rightarrow$ \$0.89) 
on \textsc{Who\&When} Hand-Crafted, an overhead of less than 3.5\% in latency as well, because of the lightweight schema. Since schema retrieval
adds only a small amount of contextual guidance to the prompt and requires no fine-tuning, the approach remains lightweight and
compatible with real-time MAS deployments.

\paragraph{Sensitivity to Similarity and Hotness Thresholds.}
CORRECT is robust to its two primary hyperparameters: the similarity threshold
$\delta$ and the schema "hotness" threshold $\theta_{\text{hot}}$. As shown in Table~\ref{tab:sensitivity_sim} and Table~\ref{tab:sensitivity:cache}, varying
$\theta_{\text{hot}}$ (which governs how frequently high-access schemata are
refined) produces only small but consistent gains on ARC (80.4$\rightarrow$81.3
as $\theta_{\text{hot}}$ increases from 0 to 0.3), while adjusting $\delta$
mainly trades off cache size against accuracy (76.5, 78.1, and 78.9 for
$\delta{=}0.6, 0.7, 0.8$ respectively). They indicate that
$\delta{=}0.7$ and refining the top 20\% most-accessed schemata provide a strong
balance, and that CORRECT’s performance is stable.

\paragraph{Decisive Interventions.} 
We further evaluate whether error localization can directly support targeted intervention to improve the fault tolerance of multi-agent systems. We compare targeted restart from the steps identified by \name and from random steps on \textsc{Who\&When} Hand-Crafted. We notice the end-to-end success rate improves from 10\% to 20\% with our method. This shows that even when exact Acc@0 is far from saturated, the predicted error step is still useful for downstream MAS troubleshooting and recovery.


\paragraph{Robustness to Retrieval Noise.}
To evaluate \name's robustness to retrieval noise, we inject
irrelevant ``hard-negative'' schemata into the top-$k$ retrieval pool while
keeping $k{=}5$ fixed. Accuracy decreases moderately as noise increases (e.g.,
Qwen-7B drops from 12.3\% to 6.9\% when all five retrieved schemata are random, yet \name consistently outperforms the baseline (3.5\%). GPT-5 shows a
similar trend: accuracy drops from 17.2\% with clean retrieval to 12.1\% when
all five schemata are random, while remaining above the baseline of 8.6\%.
These results show that \name does not rely on perfect retrieval. Schema
guidance remains useful even when multiple partially mismatched schemata are
included. Full results are available in Appendix~\ref{app:retrieval_noise}.

\paragraph{Cross-Domain Versatility.}
Beyond \textsc{Correct-Error}, we evaluate \name's transferability across
domains, supervision formats, and benchmarks
(Appendix~\ref{app:who_when_transfer}). In the error-category detection setting
of \citet{cemri2025multi}, \name improves recall from 15.7\% to 16.6\% on
ProgramDev+ChatDev and from 11.9\% to 15.8\% on AG2+GSM8K, despite using no
step-level annotations (Table~\ref{tab:category_transfer}). On
AgentErrorBench~\citep{zhu2025llm}, \name also improves GPT-5-nano tolerant
accuracy on GAIA (36\% to 46\%), ALFWorld (22\% to 35\%), and WebShop
(16\% to 24\%) (Table~\ref{tab:agenterrorbench_transfer}). These results show
that the distilled schemata transfer across datasets, task formulations, and
agentic environments.

\section{Limitations}
We acknowledge the following limitations. First, decisive error labels may be inherently subjective, even in human-crafted or human-annotated trajectories, since failures in MAS often involve multiple interacting agents and cascading effects. This ambiguity can introduce instability for future training, evaluation, or online adaptation that relies on such labels. Second, while \name shows that reusable error schemata are effective for error recognition, how to better integrate these schemata with skill systems in multi-agent frameworks remains an important direction. Finally, \name currently retrieves schemata mainly through semantic similarity, and it remains unclear whether this is the best retrieval signal, despite its robustness to retrieval noises and low costs. We leave the exploration of more structured retrieval mechanisms, potentially with a dedicated retrieval agent that reasons over schemata, agent roles, tool states, and causal dependencies, to future work.

\section{Conclusion}
We introduced \name, the first schema-guided framework that distills recurrent MAS failures into compact, reusable schemata, enabling accurate, lightweight, and training-free identification of decisive errors in new runs. Complementing this, we release \textsc{CORRECT-Error}, a large-scale, high-fidelity benchmark capturing realistic error patterns. \name significantly improves error recognition accuracy, offering a practical, generalizable path toward reliable, interpretable, and scalable MAS deployment.

\section*{Acknowledgements}
We thank the anonymous reviewers for their constructive and insightful feedback. This work was
supported in part by an Amazon Research Award and awards from NVIDIA Academic Program and Gemini Academic Program. It also utilized the Delta system at the National Center for Supercomputing Applications
(NCSA) through allocation CIS240236 from the ACCESS program.

\section*{Impact Statement}
This paper presents work whose goal is to advance the reliability and interpretability of multi-agent systems by enabling precise and efficient error recognition. As multi-agent systems are increasingly deployed in real-world settings, such as software engineering, scientific discovery, and automated decision, improving their debuggability can lead to safer, more robust, and more maintainable AI systems. By identifying decisive errors and their propagation paths, our approach will help practitioners prevent cascading failures, reduce operational costs, and support responsible system upgrades and monitoring.
\nocite{langley00}

\bibliography{example_paper}

@article{zhang2025agent,
  title={Which Agent Causes Task Failures and When? On Automated Failure Attribution of LLM Multi-Agent Systems},
  author={Zhang, Shaokun and Yin, Ming and Zhang, Jieyu and Liu, Jiale and Han, Zhiguang and Zhang, Jingyang and Li, Beibin and Wang, Chi and Wang, Huazheng and Chen, Yiran and others},
  journal={arXiv preprint arXiv:2505.00212},
  year={2025}
}

@inproceedings{kwon2023efficient,
  title={Efficient Memory Management for Large Language Model Serving with PagedAttention},
  author={Woosuk Kwon and Zhuohan Li and Siyuan Zhuang and Ying Sheng and Lianmin Zheng and Cody Hao Yu and Joseph E. Gonzalez and Hao Zhang and Ion Stoica},
  booktitle={Proceedings of the ACM SIGOPS 29th Symposium on Operating Systems Principles},
  year={2023}
}

@article{agent-prod-survey25,
      title={Measuring Agents in Production}, 
      author={Melissa Z. Pan and Negar Arabzadeh and Riccardo Cogo and Yuxuan Zhu and Alexander Xiong and Lakshya A Agrawal and Huanzhi Mao and Emma Shen and Sid Pallerla and Liana Patel and Shu Liu and Tianneng Shi and Xiaoyuan Liu and Jared Quincy Davis and Emmanuele Lacavalla and Alessandro Basile and Shuyi Yang and Paul Castro and Daniel Kang and Joseph E. Gonzalez and Koushik Sen and Dawn Song and Ion Stoica and Matei Zaharia and Marquita Ellis},
      year={2025},
      journal={arXiv preprint 2512.04123}, 
}

@article{yang2025qwen3,
  title={Qwen3 technical report},
  author={Yang, An and Li, Anfeng and Yang, Baosong and Zhang, Beichen and Hui, Binyuan and Zheng, Bo and Yu, Bowen and Gao, Chang and Huang, Chengen and Lv, Chenxu and others},
  journal={arXiv preprint arXiv:2505.09388},
  year={2025}
}

@article{qian2023chatdev,
  title={Chatdev: Communicative agents for software development},
  author={Qian, Chen and Liu, Wei and Liu, Hongzhang and Chen, Nuo and Dang, Yufan and Li, Jiahao and Yang, Cheng and Chen, Weize and Su, Yusheng and Cong, Xin and others},
  journal={arXiv preprint arXiv:2307.07924},
  year={2023}
}

@inproceedings{hong2024metagpt,
      title={Meta{GPT}: Meta Programming for A Multi-Agent Collaborative Framework},
      author={Sirui Hong and Mingchen Zhuge and Jonathan Chen and Xiawu Zheng and Yuheng Cheng and Jinlin Wang and Ceyao Zhang and Zili Wang and Steven Ka Shing Yau and Zijuan Lin and Liyang Zhou and Chenyu Ran and Lingfeng Xiao and Chenglin Wu and J{\"u}rgen Schmidhuber},
      booktitle={The Twelfth International Conference on Learning Representations},
      year={2024},
      url={https://openreview.net/forum?id=VtmBAGCN7o}
}

@article{lu2024ai,
  title={The ai scientist: Towards fully automated open-ended scientific discovery},
  author={Lu, Chris and Lu, Cong and Lange, Robert Tjarko and Foerster, Jakob and Clune, Jeff and Ha, David},
  journal={arXiv preprint arXiv:2408.06292},
  year={2024}
}

@article{zhou2023webarena,
  title={Webarena: A realistic web environment for building autonomous agents},
  author={Zhou, Shuyan and Xu, Frank F and Zhu, Hao and Zhou, Xuhui and Lo, Robert and Sridhar, Abishek and Cheng, Xianyi and Ou, Tianyue and Bisk, Yonatan and Fried, Daniel and others},
  journal={arXiv preprint arXiv:2307.13854},
  year={2023}
}

@article{fourney2024magentic,
  title={Magentic-one: A generalist multi-agent system for solving complex tasks},
  author={Fourney, Adam and Bansal, Gagan and Mozannar, Hussein and Tan, Cheng and Salinas, Eduardo and Niedtner, Friederike and Proebsting, Grace and Bassman, Griffin and Gerrits, Jack and Alber, Jacob and others},
  journal={arXiv preprint arXiv:2411.04468},
  year={2024}
}

@inproceedings{wu2024autogen,
  title={Autogen: Enabling next-gen LLM applications via multi-agent conversations},
  author={Wu, Qingyun and Bansal, Gagan and Zhang, Jieyu and Wu, Yiran and Li, Beibin and Zhu, Erkang and Jiang, Li and Zhang, Xiaoyun and Zhang, Shaokun and Liu, Jiale and others},
  booktitle={First Conference on Language Modeling},
  year={2024}
}

@inproceedings{intentnet-sigcomm,
author = {Wang, Zhaodong and Lin, Samuel and Yan, Guanqing and Ghorbani, Soudeh and Yu, Minlan and Zhou, Jiawei and Hu, Nathan and Baruah, Lopa and Peters, Sam and Kamath, Srikanth and Yang, Jerry and Zhang, Ying},
title = {Intent-Driven Network Management with Multi-Agent LLMs: The Confucius Framework},
year = {2025},
booktitle = {SIGCOMM},
}

@inproceedings{mialon2023gaia,
  title={Gaia: a benchmark for general ai assistants},
  author={Mialon, Gr{\'e}goire and Fourrier, Cl{\'e}mentine and Wolf, Thomas and LeCun, Yann and Scialom, Thomas},
  booktitle={The Twelfth International Conference on Learning Representations},
  year={2023}
}

@article{dubey2024llama,
  title={The llama 3 herd of models},
  author={Dubey, Abhimanyu and Jauhri, Abhinav and Pandey, Abhinav and Kadian, Abhishek and Al-Dahle, Ahmad and Letman, Aiesha and Mathur, Akhil and Schelten, Alan and Yang, Amy and Fan, Angela and others},
  journal={arXiv e-prints},
  pages={arXiv--2407},
  year={2024}
}

@article{cemri2025multi,
  title={Why Do Multi-Agent LLM Systems Fail?},
  author={Cemri, Mert and Pan, Melissa Z and Yang, Shuyi and Agrawal, Lakshya A and Chopra, Bhavya and Tiwari, Rishabh and Keutzer, Kurt and Parameswaran, Aditya and Klein, Dan and Ramchandran, Kannan and others},
  journal={arXiv preprint arXiv:2503.13657},
  year={2025}
}

@article{achiam2023gpt,
  title={Gpt-4 technical report},
  author={Achiam, Josh and Adler, Steven and Agarwal, Sandhini and Ahmad, Lama and Akkaya, Ilge and Aleman, Florencia Leoni and Almeida, Diogo and Altenschmidt, Janko and Altman, Sam and Anadkat, Shyamal and others},
  journal={arXiv preprint arXiv:2303.08774},
  year={2023}
}

@article{SU2024127063,
title = {RoFormer: Enhanced transformer with Rotary Position Embedding},
journal = {Neurocomputing},
volume = {568},
pages = {127063},
year = {2024},
issn = {0925-2312},
doi = {https://doi.org/10.1016/j.neucom.2023.127063},
url = {https://www.sciencedirect.com/science/article/pii/S0925231223011864},
author = {Jianlin Su and Murtadha Ahmed and Yu Lu and Shengfeng Pan and Wen Bo and Yunfeng Liu}
}

@article{peng2023yarn,
  title={Yarn: Efficient context window extension of large language models},
  author={Peng, Bowen and Quesnelle, Jeffrey and Fan, Honglu and Shippole, Enrico},
  journal={arXiv preprint arXiv:2309.00071},
  year={2023}
}

@misc{bge-m3,
      title={BGE M3-Embedding: Multi-Lingual, Multi-Functionality, Multi-Granularity Text Embeddings Through Self-Knowledge Distillation}, 
      author={Jianlv Chen and Shitao Xiao and Peitian Zhang and Kun Luo and Defu Lian and Zheng Liu},
      year={2024},
      eprint={2402.03216},
      archivePrefix={arXiv},
      primaryClass={cs.CL}
}

@article{gao2025single,
  title={Single-agent or Multi-agent Systems? Why Not Both?},
  author={Gao, Mingyan and Li, Yanzi and Liu, Banruo and Yu, Yifan and Wang, Phillip and Lin, Ching-Yu and Lai, Fan},
  journal={arXiv preprint arXiv:2505.18286},
  year={2025}
}

@inproceedings{epperson2025interactive,
  title={Interactive debugging and steering of multi-agent ai systems},
  author={Epperson, Will and Bansal, Gagan and Dibia, Victor C and Fourney, Adam and Gerrits, Jack and Zhu, Erkang and Amershi, Saleema},
  booktitle={Proceedings of the 2025 CHI Conference on Human Factors in Computing Systems},
  pages={1--15},
  year={2025}
}

@article{zhang2024reactable,
  title={ReAcTable: enhancing ReAct for table question answering},
  author={Zhang, Yunjia and Henkel, Jordan and Floratou, Avrilia and Cahoon, Joyce and Deep, Shaleen and Patel, Jignesh M},
  journal={Proceedings of the VLDB Endowment},
  volume={17},
  number={8},
  pages={1981--1994},
  year={2024},
  publisher={VLDB Endowment}
}

@article{ge2025introducing,
  title={Who is Introducing the Failure? Automatically Attributing Failures of Multi-Agent Systems via Spectrum Analysis},
  author={Ge, Yu and Xie, Linna and Li, Zhong and Pei, Yu and Zhang, Tian},
  journal={arXiv preprint arXiv:2509.13782},
  year={2025}
}

@article{zheng2023judging,
  title={Judging llm-as-a-judge with mt-bench and chatbot arena},
  author={Zheng, Lianmin and Chiang, Wei-Lin and Sheng, Ying and Zhuang, Siyuan and Wu, Zhanghao and Zhuang, Yonghao and Lin, Zi and Li, Zhuohan and Li, Dacheng and Xing, Eric and others},
  journal={Advances in neural information processing systems},
  volume={36},
  pages={46595--46623},
  year={2023}
}

@article{Yang2024Qwen25TR,
  title={Qwen2.5 Technical Report},
  author={Qwen An Yang and Baosong Yang and Beichen Zhang and Binyuan Hui and Bo Zheng and Bowen Yu and Chengyuan Li and Dayiheng Liu and Fei Huang and Guanting Dong and Haoran Wei and Huan Lin and Jian Yang and Jianhong Tu and Jianwei Zhang and Jianxin Yang and Jiaxin Yang and Jingren Zhou and Junyang Lin and Kai Dang and Keming Lu and Keqin Bao and Kexin Yang and Le Yu and Mei Li and Mingfeng Xue and Pei Zhang and Qin Zhu and Rui Men and Runji Lin and Tianhao Li and Tingyu Xia and Xingzhang Ren and Xuancheng Ren and Yang Fan and Yang Su and Yi-Chao Zhang and Yunyang Wan and Yuqi Liu and Zeyu Cui and Zhenru Zhang and Zihan Qiu and Shanghaoran Quan and Zekun Wang},
  journal={ArXiv},
  year={2024},
  volume={abs/2412.15115},
  url={https://api.semanticscholar.org/CorpusID:274859421}
}

@misc{openai2025gpt5,
  author       = {OpenAI},
  title        = {GPT-5},
  year         = {2025},
  note         = {Large language model, accessed via ChatGPT},
  howpublished = {\url{https://openai.com}},
}

@article{hurst2024gpt,
  title={Gpt-4o system card},
  author={Hurst, Aaron and Lerer, Adam and Goucher, Adam P and Perelman, Adam and Ramesh, Aditya and Clark, Aidan and Ostrow, AJ and Welihinda, Akila and Hayes, Alan and Radford, Alec and others},
  journal={arXiv preprint arXiv:2410.21276},
  year={2024}
}

@article{guo2025deepseek,
  title={Deepseek-r1: Incentivizing reasoning capability in llms via reinforcement learning},
  author={Guo, Daya and Yang, Dejian and Zhang, Haowei and Song, Junxiao and Zhang, Ruoyu and Xu, Runxin and Zhu, Qihao and Ma, Shirong and Wang, Peiyi and Bi, Xiao and others},
  journal={arXiv preprint arXiv:2501.12948},
  year={2025}
}

@article{comanici2025gemini,
  title={Gemini 2.5: Pushing the frontier with advanced reasoning, multimodality, long context, and next generation agentic capabilities},
  author={Comanici, Gheorghe and Bieber, Eric and Schaekermann, Mike and Pasupat, Ice and Sachdeva, Noveen and Dhillon, Inderjit and Blistein, Marcel and Ram, Ori and Zhang, Dan and Rosen, Evan and others},
  journal={arXiv preprint arXiv:2507.06261},
  year={2025}
}

@inproceedings{yang2018hotpotqa,
  title={{HotpotQA}: A Dataset for Diverse, Explainable Multi-hop Question Answering},
  author={Yang, Zhilin and Qi, Peng and Zhang, Saizheng and Bengio, Yoshua and Cohen, William W. and Salakhutdinov, Ruslan and Manning, Christopher D.},
  booktitle={Conference on Empirical Methods in Natural Language Processing ({EMNLP})},
  year={2018}
}

@article{trivedi2022musique,
  title={MuSiQue: Multihop Questions via Single-hop Question Composition},
  author={Trivedi, Harsh and Balasubramanian, Niranjan and Khot, Tushar and Sabharwal, Ashish},
  journal={Transactions of the Association for Computational Linguistics},
  volume={10},
  pages={539--554},
  year={2022},
  publisher={MIT Press One Broadway, 12th Floor, Cambridge, Massachusetts 02142, USA~…}
}

@inproceedings{xanh2020_2wikimultihop,
    title = "Constructing A Multi-hop {QA} Dataset for Comprehensive Evaluation of Reasoning Steps",
    author = "Ho, Xanh  and
      Duong Nguyen, Anh-Khoa  and
      Sugawara, Saku  and
      Aizawa, Akiko",
    booktitle = "Proceedings of the 28th International Conference on Computational Linguistics",
    month = dec,
    year = "2020",
    address = "Barcelona, Spain (Online)",
    publisher = "International Committee on Computational Linguistics",
    url = "https://www.aclweb.org/anthology/2020.coling-main.580",
    pages = "6609--6625",
}

@article{allenai:arc,
      author    = {Peter Clark  and Isaac Cowhey and Oren Etzioni and Tushar Khot and
                    Ashish Sabharwal and Carissa Schoenick and Oyvind Tafjord},
      title     = {Think you have Solved Question Answering? Try ARC, the AI2 Reasoning Challenge},
      journal   = {arXiv:1803.05457v1},
      year      = {2018},
}

@article{wang2024mmlu,
  title={Mmlu-pro: A more robust and challenging multi-task language understanding benchmark},
  author={Wang, Yubo and Ma, Xueguang and Zhang, Ge and Ni, Yuansheng and Chandra, Abhranil and Guo, Shiguang and Ren, Weiming and Arulraj, Aaran and He, Xuan and Jiang, Ziyan and others},
  journal={Advances in Neural Information Processing Systems},
  volume={37},
  pages={95266--95290},
  year={2024}
}

@article{lightman2023lets,
      title={Let's Verify Step by Step}, 
      author={Lightman, Hunter and Kosaraju, Vineet and Burda, Yura and Edwards, Harri and Baker, Bowen and Lee, Teddy and Leike, Jan and Schulman, John and Sutskever, Ilya and Cobbe, Karl},
      journal={arXiv preprint arXiv:2305.20050},
      year={2023}
}

@inproceedings{10.1145/3731569.3764829,
author = {Yu, Yifan and Gan, Yu and Sarda, Nikhil and Tsai, Lillian and Shen, Jiaming and Zhou, Yanqi and Krishnamurthy, Arvind and Lai, Fan and Levy, Hank and Culler, David},
title = {IC-Cache: Efficient Large Language Model Serving via In-context Caching},
year = {2025},
isbn = {9798400718700},
publisher = {Association for Computing Machinery},
address = {New York, NY, USA},
url = {https://doi.org/10.1145/3731569.3764829},
doi = {10.1145/3731569.3764829},
booktitle = {Proceedings of the ACM SIGOPS 31st Symposium on Operating Systems Principles},
pages = {375–398},
numpages = {24},
location = {Lotte Hotel World, Seoul, Republic of Korea},
series = {SOSP '25}
}

@article{peng2023instruction,
  title={Instruction tuning with gpt-4},
  author={Peng, Baolin and Li, Chunyuan and He, Pengcheng and Galley, Michel and Gao, Jianfeng},
  journal={arXiv preprint arXiv:2304.03277},
  year={2023}
}

@article{chen2025step,
  title={Step-wise Adaptive Integration of Supervised Fine-tuning and Reinforcement Learning for Task-Specific LLMs},
  author={Chen, Jack and Liu, Fazhong and Liu, Naruto and Luo, Yuhan and Qin, Erqu and Zheng, Harry and Dong, Tian and Zhu, Haojin and Meng, Yan and Wang, Xiao},
  journal={arXiv preprint arXiv:2505.13026},
  year={2025}
}

@article{fu2025srft,
  title={SRFT: A Single-Stage Method with Supervised and Reinforcement Fine-Tuning for Reasoning},
  author={Fu, Yuqian and Chen, Tinghong and Chai, Jiajun and Wang, Xihuai and Tu, Songjun and Yin, Guojun and Lin, Wei and Zhang, Qichao and Zhu, Yuanheng and Zhao, Dongbin},
  journal={arXiv preprint arXiv:2506.19767},
  year={2025}
}

@article{opsahl2024optimizing,
  title={Optimizing instructions and demonstrations for multi-stage language model programs},
  author={Opsahl-Ong, Krista and Ryan, Michael J and Purtell, Josh and Broman, David and Potts, Christopher and Zaharia, Matei and Khattab, Omar},
  journal={arXiv preprint arXiv:2406.11695},
  year={2024}
}

@article{zhu2025llm,
  title={Where llm agents fail and how they can learn from failures},
  author={Zhu, Kunlun and Liu, Zijia and Li, Bingxuan and Tian, Muxin and Yang, Yingxuan and Zhang, Jiaxun and Han, Pengrui and Xie, Qipeng and Cui, Fuyang and Zhang, Weijia and others},
  journal={arXiv preprint arXiv:2509.25370},
  year={2025}
}

@article{zheng2024llamafactory,
  title={Llamafactory: Unified efficient fine-tuning of 100+ language models},
  author={Zheng, Yaowei and Zhang, Richong and Zhang, Junhao and Ye, Yanhan and Luo, Zheyan and Feng, Zhangchi and Ma, Yongqiang},
  journal={arXiv preprint arXiv:2403.13372},
  year={2024}
}
\bibliographystyle{icml2026}

\newpage
\appendix
\onecolumn
\section{Appendix}
\subsection{Impact Statement}
\label{sec:app_reproduce}
\paragraph{Impact Statement.} This paper presents a method for improving error recognition in large language model–based multi-agent systems, with the goal of enhancing their reliability, interpretability, and maintainability. By enabling more accurate and efficient diagnosis of failure sources without additional training, our work can reduce debugging costs and support safer deployment of complex AI systems in research and real-world applications. We do not foresee significant negative societal impacts arising from this work when used responsibly.


\subsection{MAS failure trajectories are long and complex}
\label{app:failure-length}

\begin{wrapfigure}{r}{0.3\textwidth}
    \centering
    \includegraphics[width=\linewidth]{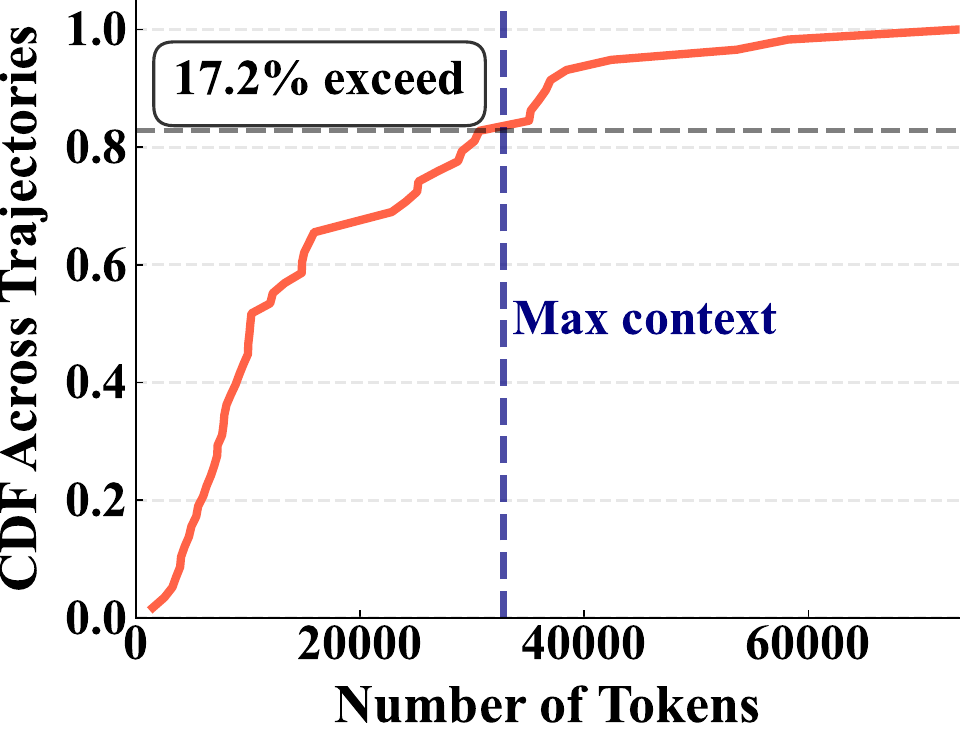}
    \caption{MAS failure traces are complex and long, often exceeding the model capacity.}
    \label{fig:token_dist}
\vspace{-1cm}
\end{wrapfigure}
We show the distribution of MAS trajectories as in~\ref{fig:token_dist}. About 17.2\% of the trajectories exceed the length limit of max content limit of Qwen models.

\subsection{Specifics of experimental settings}
\label{sec:app_exp_settings}
We implement our evaluation pipeline based on \citet{zhang2025agent}. We host open-source models using vLLM\citep{kwon2023efficient} and access GPT-series models\citep{achiam2023gpt} via the OpenAI API. To handle long contexts exceeding standard model limits for Qwen models\citep{yang2025qwen3}, we employ RoPE\citep{SU2024127063} scaling with 4× length extension using the "yarn"\citep{peng2023yarn} scaling type. To simulate realistic deployment scenarios where ground truth is 
\begin{wrapfigure}{r}{0.3\textwidth}
    \centering
    \includegraphics[width=\linewidth]{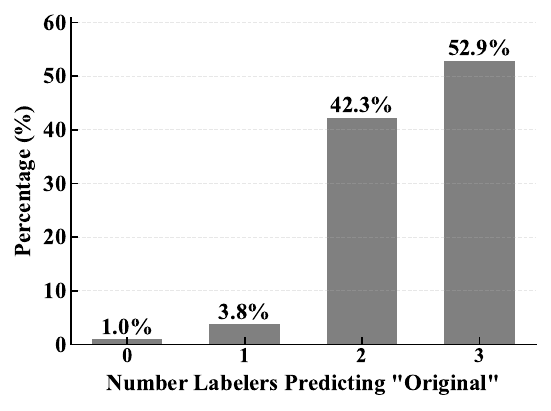}
    \caption{Percentage of human labelers to believe the trajectory is not synthesized.}
    \label{fig:app_data_human_agreement}
\vspace{-1cm}
\end{wrapfigure}
unknown, we exclude the correct answer from evaluation prompts. 
For our method, we first generate all the error schemata using GPT-5 model. We then derive a similarity mapping to assign error schemata based on the semantic embedding decoded by BAAI-BPE-M3 model\citep{bge-m3}. 

As shown in
Appendix~\ref{app:context_truncation}, \name is robust to aggressive
truncation: using 500, 1000, or 2000 characters per turn yields nearly
identical performance, indicating that long MAS logs (often 60k+ tokens) do not
pose a practical limitation for schema retrieval.

To avoid the data leakage, we mask each trajectory itself and avoid receiving its own error schema. We decide the number of error schemata from the experiments using Qwen-2.5-7b models on Hand-Crafted dataset, Algorithm-Generated dataset, and HotpotQA dataset of CORRECT-Error. 
We use the same number of error schemata across all models and all datasets in CORRECT-Error. Specifically, we use 1 error schema for all experiments on the Algorithm-Generated dataset, 10 error schemata for all experiments on the Hand-Crafted dataset, and 5 error schemata for all experiments on CORRECT-Error.

\paragraph{Fine-Tuning Baseline Details.}
Following prior work~\citep{zheng2024llamafactory,peng2023instruction},
we train the model with a standard cross-entropy loss over the assistant
responses. To avoid data leakage, SFT for the Hand-Crafted evaluation uses
training data from the Algorithm-Generated split, and vice versa.

Each training instance is formatted as:

\texttt{messages = [} \\
\texttt{\ \ \{"role": "system", "content": "You are an AI assistant specialized in analyzing multi-agent conversations to identify errors."\},} \\
\texttt{\ \ \{"role": "user", "content": prompt\},} \\
\texttt{\ \ \{"role": "assistant", "content": answer\}} \\
\texttt{]}

Only the assistant output is used as the training target.

We perform a grid search over learning rates \{1e-6, 5e-6, 1e-5, 5e-5\} and
batch sizes \{8, 16, 32\}. The selected hyperparameters for the two SFT models
are:

\begin{center}
\begin{tabular}{lcc}
\toprule
 & Learning Rate & Batch Size \\
\midrule
Hand-Crafted SFT & 5e-5 & 16 \\
Algorithm-Generated SFT & 1e-5 & 32 \\
\bottomrule
\end{tabular}
\end{center}

These models are trained on the full trajectory dataset for each split to encode
domain-specific failure patterns.

\subsection{More details of the CORRECT-Error}
\label{app:alignment}
We implemented a variant of Magentic-One~\citep{fourney2024magentic} using selector group based workflow control using AutoGen~\citep{wu2024autogen} to generate CORRECT-Error. 
Apart from the figures we showed in section~\ref{sec:dataset}, we observed strong inter-annotator consensus: 94.4\% of synthetic trajectories fooled at least two labelers, while 52.9\% were unanimously mistaken for genuine errors. We show the distribution in Figure~\ref{fig:app_data_human_agreement}

\subsection{Prompts for Offline Error Schema Generation}
\label{sec:app_offline}
We show the prompts we used for offline schema generation in Fig~\ref{fig:app_schema_create}.
\begin{figure*}
\caption{Error Schema Generation}
\label{fig:app_schema_create}

\begin{llmprompt}{Error Schema Generation}
"""Given an error analysis from a multi-agent conversation, create an error schema to help identify similar errors in the future.

Context:

Question: \{question\}

Ground Truth: \{ground\_truth\}

Error Agent: \{mistake\_agent\}

Error Step: \{mistake\_step\}

Error Reason: \{mistake\_reason\}

Conversation History:
\{chat\_content\}

Based on this error case, please create a error schema that will help IDENTIFY similar errors in future conversations. Focus primarily on recognition patterns rather than mitigation strategies. The schema should include:

1. Error Signatures:
   - What distinctive patterns or signals indicate this type of error is occurring?
   - What are the telltale signs in the agent's behavior or responses?

2. Error Context Analysis:
   - What contextual conditions typically surround this type of error?
   - What sequence of interactions tends to precede this error?

3. Detection Heuristics:
   - What specific questions can be asked to determine if this error is present?
   - What analytical framework can help identify this error pattern?
   - What key phrases or conversation patterns serve as reliable indicators?

Please format your response as a structured schema that focuses specifically on ERROR IDENTIFICATION, not on how to improve agent behavior.

Provide a concise, actionable schema in the following format:

Agent Name: \{mistake\_agent\}

Step Number: \{mistake\_step\}

Reason for Mistake: [Your analysis of why this specific error occurred and how to identify similar patterns]
"""

\end{llmprompt}
\end{figure*}

\subsection{Prompts for online schema-guided generation}
\label{sec:app_online}
We show the prompts we used for online schema-guided generation in Fig~\ref{fig:app_schema_gen}.

\begin{figure*}
\caption{Schema-guided generation}
\label{fig:app_schema_gen}
\begin{llmprompt2}{Schema-guided generation}
\footnotesize
"HOW TO USE THIS REFERENCE EXAMPLE:"

                    "This template demonstrates one type of error pattern for reference. To apply it to your analysis:"
                    
                    "1. Study the ERROR PATTERN shown: What type of mistake does this example identify?"
                    
                    "2. Use this as reference to analyze YOUR conversation:"
                    
                    "   • Read through your conversation systematically (Step 0, Step 1, Step 2...)"
                    
                    "   • At each step, ask: 'Is there an error here, and does it match this pattern or a different one?'"
                    
                    "   • The error in your case may follow the same pattern or be completely different"
                    
                    "3. Remember this is just a reference example:"
                    
                    "   • Your error may occur at any step number"
                    
                    "   • Your error may be a different type entirely"
                    
                    "   • Use this template to help you recognize what errors look like, not to assume your error matches"

\end{llmprompt2}

\end{figure*}





\subsection{Comparison to Human-Curated Taxonomies}
\label{app:mast_comparison}

We further compare and contrast the error schemata generated by CORRECT with
human-curated taxonomies in MAST:

\textbf{Error Schemata vs. Static Human Taxonomies}
Human-curated taxonomies face several limitations:

\begin{denseitemize}

\item \emph{Lack of granularity.} Human-defined categories (e.g., ``wrong tool
use,'' ``missing precondition'') operate at coarse conceptual levels and do
not provide the fine-grained, step-specific patterns required for localizing
errors within trajectories.

\item \emph{High manual cost and slow iteration.} Curating and maintaining a taxonomy
requires extensive domain expertise. As models, tools, and task distributions
shift, these taxonomies quickly become outdated.

\item  \emph{Limited adaptability under distribution drift.} Static taxonomies cannot
capture new, emerging failure modes as the environment or agents evolve.

\end{denseitemize}

\textbf{Advantages of CORRECT.}
CORRECT automatically distills fine-grained, step-level schemata from observed
trajectories and updates them asynchronously without adding online inference
overhead. This enables CORRECT to track evolving failure modes and maintain
accuracy under distribution drift—capabilities that static human-curated
taxonomies cannot provide in large-scale deployments.

\subsection{Human Verification, Automated Validation, and Scalability}
\label{app:validation}

Human involvement during schema construction can introduce overhead, but CORRECT
is not dependent on manual supervision. In practical deployments (e.g.,
conversational systems such as ChatGPT or Gemini), user feedback is already
collected passively and can naturally serve as weak supervision without incurring
additional annotation costs. Moreover, each schema is reused across many queries,
so even a small number of validated schemata provides substantial amortized
benefit: as shown in Figure~8, only 58 schemata are sufficient to improve
HotpotQA detection accuracy from 14.7 to 35.4.

When human input is unavailable, CORRECT can instead employ LLM-as-a-judge with
confidence filtering to automatically validate successful trajectories before
inserting new schemata into the cache. This aligns with recent efforts such as
EcoAssistant, which leverage human-verified or automatically filtered caches to
improve long-horizon reasoning. Exploring more automated or semi-supervised
schema-validation pipelines—while preserving interpretability and reliability—is
a promising direction for future work.

\subsection{Prompt for Natural Error Injection}
\label{app:error_injection_prompt}
\begin{figure*}
\begin{llmprompt}{Error injection}
\footnotesize
\texttt{You are analyzing an error pattern to create an injection plan.} \\[4pt]

\texttt{THOUGHT TEMPLATE (Error Detection Guidelines):} \\
\texttt{\{thought\_template\}} \\[4pt]

\texttt{ERROR INFORMATION:} \\
\texttt{- Agent: \{mistake\_agent\}} \\
\texttt{- Step: \{mistake\_step\}} \\
\texttt{- Reason: \{mistake\_reason\}} \\[4pt]

\texttt{ERROR TRAJECTORY:} \\
\texttt{\{error\_formatted\}} \\[4pt]

\texttt{CORRECT TRAJECTORY:} \\
\texttt{\{correct\_formatted\}} \\[4pt]

\texttt{CRITICAL REQUIREMENTS:} \\
\texttt{1. Injection point must be semantically similar to the original error.} \\
\texttt{2. Error must flow naturally from that point.} \\
\texttt{3. Consider cascade effects on all later steps.} \\
\texttt{4. For consecutive CodeExecutor steps, modify only the final step's output.} \\[4pt]

\texttt{STEP-BY-STEP ANALYSIS REQUIRED:} \\
\texttt{Step 1: Analyze the original error and identify what went wrong.} \\
\texttt{Step 2: Map to semantically similar steps in the correct trajectory.} \\
\texttt{Step 3: Select the proper injection point.} \\
\texttt{Step 4: Describe how the error should be replicated.} \\
\texttt{Step 5: Assess how the injected error propagates.} \\[4pt]

\texttt{VERIFICATION CHECKLIST:} \\
\texttt{- Semantically similar injection point?} \\
\texttt{- Natural error flow?} \\
\texttt{- Cascade effects considered?} \\
\texttt{- Final CodeExecutor step modified if consecutive?} \\
\texttt{- Realistic and agent-consistent error?} \\[4pt]

\texttt{Output format:} \\
\texttt{<injection\_step>[...]</injection\_step>} \\
\texttt{<injection\_agent>[...]</injection\_agent>} \\
\texttt{<error\_pattern>[...]</error\_pattern>} \\
\texttt{<injection\_strategy>[...]</injection\_strategy>} \\
\texttt{<expected\_impact>[...]</expected\_impact>} \\

\end{llmprompt}
\end{figure*}

\begin{figure*}

\begin{llmprompt}
\footnotesize
\texttt{You are tasked with modifying a message to inject a specific error.} \\
\texttt{Follow the steps below carefully.} \\[4pt]

\texttt{ORIGINAL MESSAGE:} \\
\texttt{Agent: \{agent\}} \\
\texttt{Content: \{original\_content\}} \\[6pt]

\texttt{INJECTION PLAN:} \\
\texttt{Error Pattern: \{injection\_plan['error\_pattern']\}} \\
\texttt{Strategy: \{injection\_plan['error\_strategy']\}} \\
\texttt{Expected Impact: \{injection\_plan['expected\_impact']\}} \\[6pt]

\texttt{INSTRUCTIONS: Complete each section below in order. Show your
thinking process.} \\[6pt]

\texttt{<STEP1\_ANALYSIS>} \\
\texttt{- What is the agent's formatting style?} \\
\texttt{- What was the agent trying to communicate?} \\
\texttt{- What is the context of this message?} \\
\texttt{</STEP1\_ANALYSIS>} \\[6pt]

\texttt{<STEP2\_ERROR\_UNDERSTANDING>} \\
\texttt{- What specific error am I injecting? \{injection\_plan['error\_strategy']\}} \\
\texttt{- Why would this error naturally occur?} \\
\texttt{- How does this relate to the error pattern
\{injection\_plan['error\_pattern']\}?} \\
\texttt{</STEP2\_ERROR\_UNDERSTANDING>} \\[6pt]

\texttt{<STEP3\_MODIFICATION\_PLAN>} \\
\texttt{- What exact change will I make?} \\
\texttt{- How will I maintain the agent's style?} \\
\texttt{- Why is this change realistic?} \\
\texttt{</STEP3\_MODIFICATION\_PLAN>} \\[6pt]

\texttt{<STEP4\_VERIFICATION>} \\
\texttt{$\square$ Maintains \{agent\}'s style and format?} \\
\texttt{$\square$ Implements strategy exactly?} \\
\texttt{$\square$ Believable to the agent?} \\
\texttt{$\square$ Causes \{injection\_plan['expected\_impact']\}?} \\
\texttt{$\square$ Within agent capabilities?} \\
\texttt{$\square$ Realistic error?} \\
\texttt{$\square$ Leads to incorrect final answer?} \\
\texttt{</STEP4\_VERIFICATION>} \\[6pt]

\texttt{<MODIFIED\_CONTENT>} \\
\texttt{[Put ONLY the modified content here, exactly as the agent would output it]} \\
\texttt{</MODIFIED\_CONTENT>} \\
\end{llmprompt}
\end{figure*}

\subsection{Robustness to Context-Length Truncation}
\label{app:context_truncation}

We use the BGE-M3 encoder (context window 8192) to embed multi-agent trajectories.
Because raw MAS logs can exceed 60k--70k tokens, we truncate each turn to a fixed
character budget before embedding. Table~\ref{tab:context_trunc} shows that
CORRECT is highly robust to this truncation: using 500, 1000, or 2000 characters
per turn yields nearly identical performance.

\begin{table}[h!]
\centering
\caption{Effect of per-turn truncation length on schema retrieval accuracy.}
\label{tab:context_trunc}

\begin{tabular}{lccc}
\hline
Dataset & 500 chars & 1000 chars & 2000 chars \\
\hline
Hand-Crafted        & 12.2 & 12.2 & 12.2 \\
Algorithm-Generated & 19.8 & 19.8 & 19.2 \\
\hline
\end{tabular}

\end{table}

These results show that CORRECT’s retrieval mechanism remains stable even under
aggressive truncation, and long-context MAS logs do not hinder schema matching
in practice.

\subsection{Failure Cases: Multi-Error Trajectories}
\label{app:multi_error}

CORRECT may misfire when a trajectory contains multiple errors and a later,
higher-salience failure dominates the schema match. Because CORRECT follows the
decisiveness definition of~\citep{zhang2025agent}, it is designed to identify the
earliest decisive error; however, when downstream errors exhibit stronger
structural signals, the retrieved schema may align more closely with these
later steps.

\textbf{Example.}
In the example below, the human-labeled decisive error occurs at Step~4:
\vspace{-.3cm}
\begin{quote}
\emph{“WebSurfer failed to locate the specific volume in the University of
Leicester paper due to incomplete data retrieval and insufficient PDF analysis.”}
\end{quote}
\vspace{-.3cm}

CORRECT instead predicts Step~12 as the decisive error:
\vspace{-.3cm}
\begin{quote}
\emph{“The agent remained on the search-results page instead of navigating into
the DOI page containing the required endnote information.”}
\end{quote}
\vspace{-.3cm}

\textbf{Schema Match.}
The retrieved schema (abridged) emphasizes:
\begin{denseitemize}
\item \textbf{Error signature:} incomplete or insufficient search criteria,
\item \textbf{Context:} tasks requiring precise filtering of external data
sources,
\item \textbf{Heuristic:} “Is the agent’s search action complete and accurate?”
\end{denseitemize}

Both Step~4 and Step~12 satisfy these conditions, but Step~12 provides a more
overt instance of incorrect search behavior, making it a stronger surface-level
match for the schema.

\textbf{Discussion.}
Such multi-error trajectories represent the primary category where CORRECT may
deviate from human-annotated earliest-error labels. In practice, they are
uncommon, but they highlight an inherent challenge of schema-guided detection
when structurally similar failures occur at multiple points in a trajectory.








\subsection{Sensitivity to similarity and hotness threshold}
\label{app:sensitivity}
\begin{table}[h!]
\footnotesize
\centering
\caption{Sensitivity to the schema-refinement threshold $\theta_{\text{hot}}$.}
\label{tab:sensitivity_sim}
\begin{tabular}{lcccc}
\hline
$\theta_{\text{hot}}$ & 0.0 & 0.1 & 0.2 & 0.3 \\
\hline
ARC Accuracy & 80.4 & 80.6 & 81.2 & 81.3 \\
\hline
\end{tabular}
\end{table}

\paragraph{Effect of $\delta$.}
Lowering the similarity threshold reduces cache size but slightly harms
accuracy, whereas $\delta=0.7$ offers a strong balance:

\begin{table}[h!]
\footnotesize
\centering
\caption{Sensitivity to the similarity threshold $\delta$.}
\label{tab:sensitivity:cache}
\begin{tabular}{lccc}
\hline
$\delta$ & 0.6 & 0.7 & 0.8 \\
\hline
ARC Accuracy & 76.5 & 78.1 & 78.9 \\
\hline
\end{tabular}

\end{table}

\subsection{Transferability to Alternative Domains and Supervision Formats}
\label{app:who_when_transfer}

We further evaluate CORRECT’s cross-domain and cross-formulation generalization
under settings that differ substantially from our primary step-level error
localization task.

\textbf{Transfer to Error-Category Detection.}
The dataset introduced in~\citep{cemri2025multi} provides error \emph{categories} but
does not include step-level annotations. This differs from our formulation, where
CORRECT predicts the earliest decisive step in a trajectory. Because of this
structural mismatch, a direct step-level comparison is not possible.
To evaluate cross-formulation transferability, we adapt CORRECT by providing only
the retrieved schemata and prompting the model to output an error \emph{type}
instead of a step index. 
Using Qwen-7B, CORRECT improves recall on both released datasets:

\begin{table}[h!]
\centering
\footnotesize
\caption{CORRECT adapted to the error-category detection setting of \citep{cemri2025multi}.}
\label{tab:category_transfer}

\begin{tabular}{lcc}
\hline
Dataset & Baseline & CORRECT \\
\hline
ProgramDev + ChatDev & 15.7 & 16.6 \\
AG2 + GSM8K          & 11.9 & 15.8 \\
\hline
\end{tabular}

\end{table}

Despite lacking step-level supervision, CORRECT retains consistent gains,
indicating that its schemata encode transferable semantic error patterns that
remain effective under coarser labeling regimes.

\textbf{Transfer to AgentErrorBench.}
We further evaluate CORRECT on AgentErrorBench, which contains heterogeneous
agentic environments spanning embodied navigation, web interaction, and
long-horizon planning. This benchmark differs substantially from CORRECT-Error in
both task structure and failure modes. 
We conduct experiments using GPT-5-nano and report tolerant accuracy within
$\pm1$ step. Results are summarized in Table~\ref{tab:agenterrorbench_transfer}.

\begin{table}[h!]
\centering
\footnotesize
\caption{Cross-domain transfer results on AgentErrorBench using GPT-5-nano.}
\label{tab:agenterrorbench_transfer}

\begin{tabular}{lccc}
\hline
Dataset & Baseline & CORRECT & Improvement \\
\hline
GAIA     & 36\% & 46\% & +10\% \\
ALFWorld & 22\% & 35\% & +13\% \\
WebShop  & 16\% & 24\% & +8\%  \\
\hline
\end{tabular}

\end{table}

CORRECT achieves consistent improvements across all evaluated environments,
with the largest gains observed on ALFWorld, which requires long-horizon
planning and precise tool usage. These results demonstrate that CORRECT’s
distilled schemata generalize effectively to previously unseen domains and
interaction modalities.

\subsection{Robustness to Retrieval Noise}
\label{app:retrieval_noise}

To assess CORRECT’s resilience to mismatched schema retrieval, we inject
random, irrelevant schemata (“hard negatives’’) into the retrieval pool while
keeping the total number of retrieved schemata fixed at $k{=}5$. The tables
below report performance as the number of injected random schemata increases.

\begin{table}[h]
\footnotesize
\centering
\caption{Impact of random (irrelevant) schemata injected into the retrieval
pool (Qwen-7B).}

\begin{tabular}{lcccccc}
\hline
Model & Baseline & CORRECT & +1 Rand & +3 Rand & +4 Rand & +5 Rand \\
\hline
Qwen-7B & 3.5 & 12.3 & 10.7 & 10.7 & 10.7 & 6.9 \\
\hline
\end{tabular}

\end{table}

\begin{table}[h!]
\centering
\footnotesize
\caption{Impact of retrieval noise on GPT-5.}
\begin{tabular}{lcccccc}
\hline
Model & Baseline & CORRECT & +1 Rand & +3 Rand & +4 Rand & +5 Rand \\
\hline
GPT-5 & 8.6 & 17.2 & 13.8 & 15.5 & 13.8 & 12.1 \\
\hline
\end{tabular}

\end{table}

Even with multiple hard negatives included, CORRECT consistently improves over
the baseline, demonstrating strong robustness to retrieval noise and confirming
that schema-guided reasoning does not depend on perfect semantic matching.

\end{document}